\documentclass[twocolumn,11pt]{article}
\usepackage{graphicx,amssymb,url}
\begin{document}
\newcommand{\beq}{\begin{equation}}
\newcommand{\eeq}{\end{equation}}
\newcommand{\beqa}{\begin{eqnarray}}
\newcommand{\eeqa}{\end{eqnarray}}
\newcommand{\bs}[2]{{#1}_{\scriptscriptstyle {\rm #2}}}
\newcommand{\Hub}{\bs{H}{0}}
\newcommand{\hub}{\bs{h}{0}}
\newcommand{\age}{\bs{t}{0}}
\newcommand{\Otot}{\bs{\Omega}{TOT}}
\newcommand{\Ototo}{\bs{\Omega}{TOT,0}}
\newcommand{\Omat}{\bs{\Omega}{M}}
\newcommand{\Omato}{\bs{\Omega}{M,0}}
\newcommand{\Obar}{\bs{\Omega}{BAR}}
\newcommand{\Olum}{\bs{\Omega}{LUM}}
\newcommand{\Obdm}{\bs{\Omega}{BDM}}
\newcommand{\Ocdm}{\bs{\Omega}{CDM}}
\newcommand{\Onu}{\Omega_{\nu}}
\newcommand{\Ogam}{\Omega_{\gamma}}
\newcommand{\Orad}{\bs{\Omega}{R}}
\newcommand{\Orado}{\bs{\Omega}{R,0}}
\newcommand{\Olam}{\bs{\Omega}{\Lambda}}
\newcommand{\Olamo}{\bs{\Omega}{\Lambda,0}}
\newcommand{\Oein}{\bs{\Omega}{\Lambda,E}}
\newcommand{\rlam}{\bs{\rho}{\Lambda}}
\newcommand{\rcrit}{\bs{\rho}{crit}}
\newcommand{\rcrito}{\bs{\rho}{crit,0}}
\newcommand{\MLunits}{M_{\odot}/L_{\odot}}
\newcommand{\barf}{\bs{M}{BAR}/\bs{M}{TOT}}
\newcommand{\Lbar}{$\Lambda$bar}

\begin{center}
\fbox{ \begin{minipage}{82mm} 
\begin{center}
\begin{minipage}{79mm}
\vspace{3mm}
\parbox{77mm}
{\LARGE {\bf Problems of modern cosmology:
   How dominant is the vacuum?}}
\vspace{7mm}
\\
James Overduin,\footnote{Email: overduin@astro.uni-bonn.de}
Wolfgang Priester\footnote{Email: priester@astro.uni-bonn.de} \\
\\
Institut f\"ur Astrophysik und Extraterrestrische Forschung,
Universit\"at Bonn, Auf dem H\"ugel 71, D-53121 Bonn, Germany \\
\\
\footnotesize{It would be hard to find a cosmologist today who does not 
believe that the vast bulk of the Universe (ninety-five percent or more)
is hidden from our eyes.  We review the evidence for this remarkable consensus,
and for the latest proposal, that the mysterious dark matter consists
of as many as {\em four separate ingredients}: baryons, massive neutrinos,
new ``exotic'' dark matter particles, and vacuum 
energy, also known as the cosmological constant ($\Lambda$).
Of these, only baryons fit within standard theoretical physics; the 
others, if their existence is confirmed, will mean rewriting textbooks.
Fresh experimental evidence has recently appeared for and against all 
four components, so that the subject is in a state of turmoil and
excitement.  The past few years in particular have seen the fourth (vacuum)
component come into new prominence, largely at the expense of the third 
(exotic dark matter).  We conclude our review by exploring the 
possibility that the energy density of the vacuum is in fact so dominant
as to leave little room for significant amounts of exotic dark matter.}
\vspace{4mm}
\end{minipage} 
\end{center}
\end{minipage}}
\end{center}

\section{The Four Elements of \\ Modern Cosmology} \label{sec:int}

\begin{verse}{\footnotesize {\em 
   Hear first the four roots of all things: bright Zeus \\
   (fire), life-giving Hera (air), Aidoneus (earth) and \\ 
   Nestis (water), who moistens the springs of mor- \\
   tals with her tears.} \\
   - Empedokles, Fragments, c.~450 B.C.}
\end{verse}

Like the philosophers of antiquity, modern cosmologists have divided 
the physical world into four different realms, each characterized by
its own length scale and dominated by an increasingly rarefied species of 
invisible matter which, while not seen directly, is inferred to exist
from its gravitational influence on visible matter as well as the 
geometry of the Universe.  That which is seen --- mostly luminous hot gas,
stars and galaxies --- has a density (denoted $\Olum$) hundreds of times
smaller than the estimated total density ($\Otot$) of matter and energy
in the Universe.  Complementing it is a considerably larger amount of
so-called {\em baryonic dark matter\/} (of density $\Obdm$), which has 
a similar composition but does not shine.  Even together, however, these
two ingredients (with combined density $\Obar\equiv\Olum+\Obdm$) make up
less than five percent of the total density.  Baryonic (``ordinary'') matter
is thus relegated to a minor role in the cosmic scheme, a development 
which has rightly been seen as a ``second Copernican revolution,'' and 
one which lends a double meaning to the identification of baryons with
``earth,'' the first element of the new cosmology (Fig.~1).

\begin{figure}[t!]
\begin{center}
\includegraphics[width=77mm]{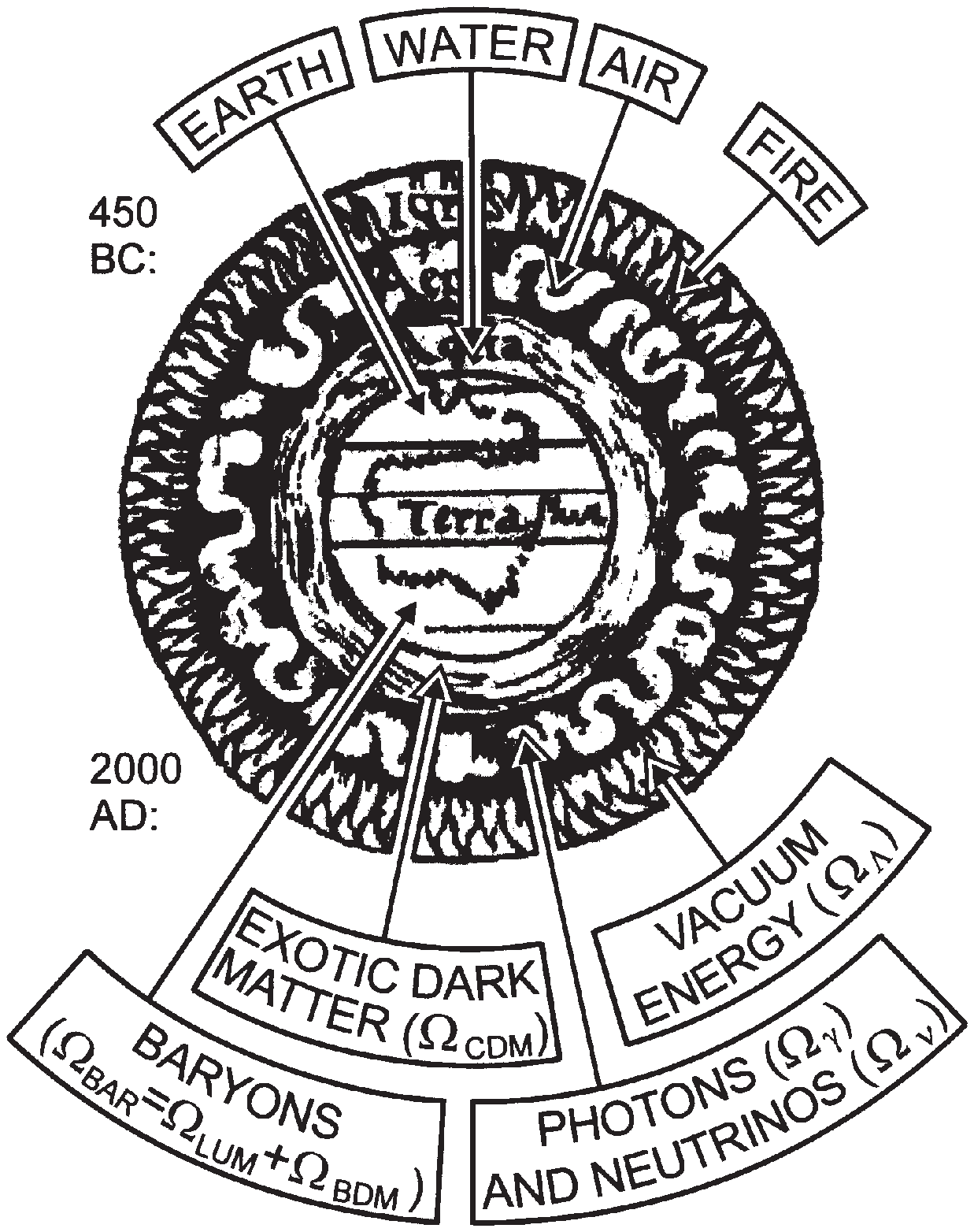}
\end{center}
\footnotesize{Fig.~1.  The ``four elements'' of modern cosmology
   (adapted from a figure in a 1519 edition of Aristotle's
   {\em Libri de Caelo\/})}
\label{4els}
\end{figure}

Observations of clumps of baryonic matter on scales larger than the
solar system (galaxies, clusters of galaxies) have led many to postulate the
existence of an additional form of matter: the ``exotic'' {\em cold dark
matter\/} (CDM) particle, whose collective density $\Ocdm$ would far exceed 
that of the baryons.  Theoretical particle physics provides several 
plausible candidates, and it could even be that this component of the
cosmic fluid --- identified with ``water'' in Fig.~1 --- itself
has several ingredients.  However, despite intensive searches, none of 
these candidate particles has yet been observed.

On still larger scales --- those relevant to the formation of galaxy
clusters in the early universe --- a different form of dark matter
may be operative: the {\em neutrino\/}.  That this particle exists is
unquestioned; but the extent of its contribution ($\Onu$) to the density 
of the Universe is not yet clear.  If the neutrino has a small (or zero) 
rest mass, then it is always relativistic and can be treated for 
dynamical purposes like the photon.  (We have therefore classified 
it together with light as the ``air'' of the new world view, Fig.~1.)
In this case, neutrino contributions can be combined with
those of photons ($\Ogam$) to give the total {\em radiation density\/} of
the Universe, $\Orad=\Onu+\Ogam$.  This is known to be insignificant at 
present.  If, on the other hand, neutrinos are sufficiently massive,
then they are no longer relativistic on average, and belong instead 
(with baryonic and exotic dark matter) under the category of
dust-like (zero-pressure) matter, with total {\em matter density\/} 
$\Omat=\Obar+\Ocdm+\Onu$.  Only in the latter case do neutrinos play 
a significant dynamical role in the present Universe.  Recent experimental
evidence has been taken by many to support this second scenario, 
implying (in most models) a collective neutrino density well below
that of baryonic matter, but (in others) a density possibly
rivalling that attributed to exotic matter.

Influential only over the largest of scales --- the cosmological horizon 
--- is the outermost species of invisible matter: the {\em vacuum
energy\/} (also known by such names as dark energy, quintessence, $x$-matter,
the zero-point field, and the cosmological constant $\Lambda$).  This 
profusion of nomenclature betrays the fact that there is at present no 
consensus as to where vacuum energy originates, or how to calculate its 
energy density ($\Olam$) from first principles.  Existing theoretical 
estimates of this latter quantity range over some 122 orders 
of magnitude, prompting most cosmologists until very recently to disregard 
it altogether.  New observations of distant supernovae, however, suggest
that the vacuum not only gravitates, but that its effective density exceeds 
that of all the other forms of matter put together.  Thus the 
``standard model'' of cosmology has evolved (in the space of 
three years) from one in which $\Olam=0$ in practice (and is as large 
as $\sim10^{122}$ in theory), to one in which there is widespread 
agreement that $\Olam$ must be of order {\em unity\/} --- indicating 
to us that cosmology, far from being a ``solved problem'' as reported 
by some authors, remains in a far-from-settled state.  We have identified
vacuum energy with ``fire'' in Fig.~1.

Four kinds of invisible matter: dark baryons, exotic particles, neutrinos
and vacuum energy --- three of which imply new physics, and any one of 
which (with the possible exception of the neutrino) outweighs the entirety
of the visible Universe.  Are they all necessary?  Our purpose in this
review will be to re-examine the steps leading up to this conclusion.
We begin with the relevant cosmological equations in \S\ref{sec:cos}.
The ``four elements'' are then introduced in turn: dark baryons 
(\S\ref{sec:bar}), exotic matter (\S\ref{sec:exo}), neutrinos 
(\S\ref{sec:nu}) and the newly-popular vacuum energy (\S\ref{sec:vac}).
In \S\ref{sec:bnp} we consider the hypothesis (advanced by one of the 
authors for ten years) that vacuum energy is in fact so dominant 
that there is no need for an exotic component to the dark matter.
Conclusions are summarized in \S\ref{sec:con}.

\section{Dark Matter and the \\ Evolution of the Universe} \label{sec:cos}

The relevant equations here are Einstein's field equations of general
relativity:\footnote{Alternative theories of gravity can also be constructed
   in such a way to remove the need for large amounts of dark matter 
   \cite{Kin01,McG01}; we do not consider these here.}
\beq
{\cal R}_{\mu\nu} - \frac{1}{2} \, {\cal R} \, g_{\mu\nu} - \Lambda \,
   g_{\mu\nu} = \frac{8\pi G}{c^{\,4}}\,{\cal T}_{\mu\nu} \; ,
\label{EFEs}
\eeq
where the terms on the left-hand side describe the geometrical structure
of spacetime, while those on the right-hand side describe its matter and
and radiation content.  ${\cal R}_{\mu\nu}$ and ${\cal R}$ are the
Ricci tensor and curvature scalar, $g_{\mu\nu}$ is the metric
tensor, and ${\cal T}_{\mu\nu}$ is the energy-momentum tensor;
$G,c$ and $\Lambda$ are all constants of nature.  About the value
of $\Lambda$, in particular, we will say more in \S\ref{sec:vac}.

The densities of matter ($\Omat$) and radiation ($\Orad$) are stored in
${\cal T}_{\mu\nu}$, while that of the vacuum ($\Olam$) is a function of
$\Lambda$.\footnote{Densities throughout this review will be written
   in the form of the dimensionless {\em density parameter\/} $\Omega$, 
   which is just the ratio of physical density ($\rho$) to the 
   {\em critical density\/} $\rcrit(t)\equiv 3H^2(t)/8\pi G$.
   This latter quantity depends on the Hubble parameter,
   Eq.~(\ref{modFE}), and takes the value $\rcrito=3\Hub^2/8\pi G$
   at the present time.  If $\Hub$ lies in the range 
   $70-90$~km~s$^{-1}$~Mpc$^{-1}$ (\S\ref{sec:bar}), then 
   $\rcrito$ is equivalent to between 5.5 and 9.1 protons
   per cubic meter.}
Assuming that these three components are distributed 
homogeneously and isotropically on large scales, and that they do not
exchange energy at a significant rate, Eqs.~(\ref{EFEs}) reduce to a 
pair of differential equations in the cosmological scale factor $R$ and
its time derivatives, including the Hubble parameter $H \equiv \dot{R}/R$ 
(the expansion rate of the Universe).  The equation for $H$ is
\beqa
[H(z)/\Hub]^2 \!\!\! & = & \!\!\! \Omato (1+z)^3 + \Orado (1+z)^4 +
   \Olamo \nonumber \\ 
   & & \!\!\! - \,(\Ototo - 1)(1+z)^2 \; ,
\label{modFE}
\eeqa
where $z \equiv (R/\bs{R}{0})^{-1}-1$ is the cosmological redshift.
The subscript ``0'' here (and throughout our review) denotes quantities 
measured at the {\em present time\/}; i.e., at redshift $z=0$.  These 
are constants, and must be carefully distinguished from functions of
time (or redshift) such as $\Omat,\Orad$ and $\Olam$.
The constant $\Ototo\equiv\Omato+\Orado+\Olamo$ is of particular interest,
because it separates spatially spherical models from hyperbolic ones.
If $\Ototo>1$, then the Universe is closed and finite in extent.
Conversely, if $\Ototo<1$, then it is open and infinite in extent.
And, if $\Ototo=1$ exactly, then we live in an infinite Universe which is 
spatially {\em flat\/} (``Euclidean'').\footnote{Strictly speaking, it is 
   also possible to obtain flat and hyperbolic solutions which are finite,
   by suitably ``identifying'' different points and adopting a nontrivial
   topology \cite{Lum98}; we do not pursue this possibility here.}
To determine the shape of the 
homogeneous and isotropic world has been a prime goal of cosmologists
since expansion was discovered.

Eq.~(\ref{modFE}) already tells us a great deal about the evolution of
the Universe.  The first term on the right-hand side, $\Omato (1+z)^3$, 
shows that matter acts to increase the expansion rate $H(z)$ as one goes 
to higher $z$ --- that is, to {\em slow down\/} the expansion with time.
This is the braking effect of matter's gravitational self-attraction.

The second term, $\Orado (1+z)^4$, shows that radiation has the same 
effect, but with a stronger dependence on redshift.  This means that, 
as one moves backward in time, photons (and relativistic particles) become 
increasingly important compared to pressureless matter.  In fact, the 
dynamics of the early Universe (at redshifts above $z\gtrsim 1000$) must
have been completely dominated by them.  At present, however, the total
density $\Orado$ of ``radiationlike matter'' (including both photons and
relativistic neutrinos) is several orders of magnitude
below that of nonrelativistic matter.\footnote{This is inferred,
   not only from measurements such as those of the {\sc Cobe} satellite, but 
   also from the fact that a too-high density of relativistic matter would 
   have slowed expansion so much that the Universe could not have lived 
   long enough to produce the oldest stars we see.}
Since we are largely concerned in this review with redshifts less than ten, 
we will drop the radiation term in Eq.~(\ref{modFE}) from this point onward.

The third term, $\Olamo$, is independent of redshift, which means that its
influence is not diluted with time.  Vacuum energy will therefore eventually
come to dominate the dynamics of the Universe in any model with $\Lambda >0$.
In the limit $t \rightarrow\infty$, in fact, the other terms drop out, and
we find that the density of the vacuum may be expressed as
$\Olamo = (H_{\infty}/\Hub)^2$, or (since 
$\Olamo\equiv\Lambda c^2/3\Hub^2$)
\beq
\Lambda c^2 = 3 H_{\infty}^2 \; ,
\label{magic}
\eeq
where $H_{\infty}$ is the limiting value of the Hubble parameter as
$t \rightarrow\infty$ (assuming that this latter quantity exists; 
i.e., that the Universe does not recollapse in the future).
This provides a little-discussed connection between $\Lambda$
(an apparent constant of nature) and the asymptotic expansion rate
(a dynamical parameter).  If $\Lambda >0$, and if we are living at
sufficiently late times, then Eq.~(\ref{magic}) immediately 
{\em predicts\/} that we will measure $\Olamo\sim1$.

The fourth term in Eq.~(\ref{modFE}), finally, shows that an excess
of $\Ototo$ over one (i.e., a positive curvature) acts to offset
the contribution of the first three terms to the expansion rate, while
a deficit (i.e., a negative curvature) enhances them.  Open models,
in other words, expand more quickly at any given redshift $z$
(and therefore last longer) than closed ones.
This curvature term, however, goes only as $(1+z)^2$, which means that
its importance drops off relative to the matter and radiation terms 
at early times, and becomes negligible compared with that of the 
vacuum term at late ones.

As recently as the 1980s, many cosmologists were persuaded that 
Eq.~(\ref{modFE}) could be substantially simplified, not only by neglecting
the second (radiation) term, but also the third (vacuum) and fourth
(curvature) terms on the right-hand side. This appeared 
reasonable at the time, for four principal reasons.  First, these terms 
differ sharply from each other (and from the first term) in their 
dependence on redshift $z$, and the probability that we should happen
to find ourselves in an era when all four terms have similar values 
would seem {\em a priori\/} very remote.  By this ``Dicke coincidence'' 
argument, it was felt that only one term ought to dominate at any 
given time \cite{Pee93}.  Second, the vacuum term in particular was
avoided for historical and theoretical reasons (to be discussed in 
\S\ref{sec:vac}).  Third, a period of cosmic {\em inflation\/} was widely 
asserted to have driven $\Otot(t)$ to exactly unity in the early 
universe.\footnote{That this is not necessarily so has been shown by several
   authors \cite{Ell88,Hub91}.  A suitable inflationary phase may be grafted
   onto models with $\Ototo\neq 1$.  The probability of finding oneself in 
   such a universe depends, not just on the raw amount of inflation, but on 
   factors such as the elapsed time and the distribution of initial 
   conditions (such as phase transitions \cite{Blo91}) which preceded
   the inflationary epoch.}
And finally, this ``standard Einstein-de~Sitter'' (EdS) model was 
favored on grounds of simplicity.  These arguments are no longer
valid today.  We are justified in neglecting the radiation term,
and {\em only\/} the radiation term in Eq.~(\ref{modFE}), leaving
\beqa
H^2(z) \!\!\! & = & \!\!\! \Hub^2 \left[ \Omato (1+z)^3 + 
   \Olamo \right. \nonumber \\ 
   & & \;\;\;\; -\left. (\Ototo - 1)(1+z)^2 \right] \; .
\label{FE}
\eeqa
This is the modern version of what is usually called {\em Friedmann's
equation\/} in cosmology.  It may be integrated numerically for the
cosmological scale factor $R(t)$ as a function of time.

\begin{figure}[t!]
\begin{center}
\includegraphics[width=85mm]{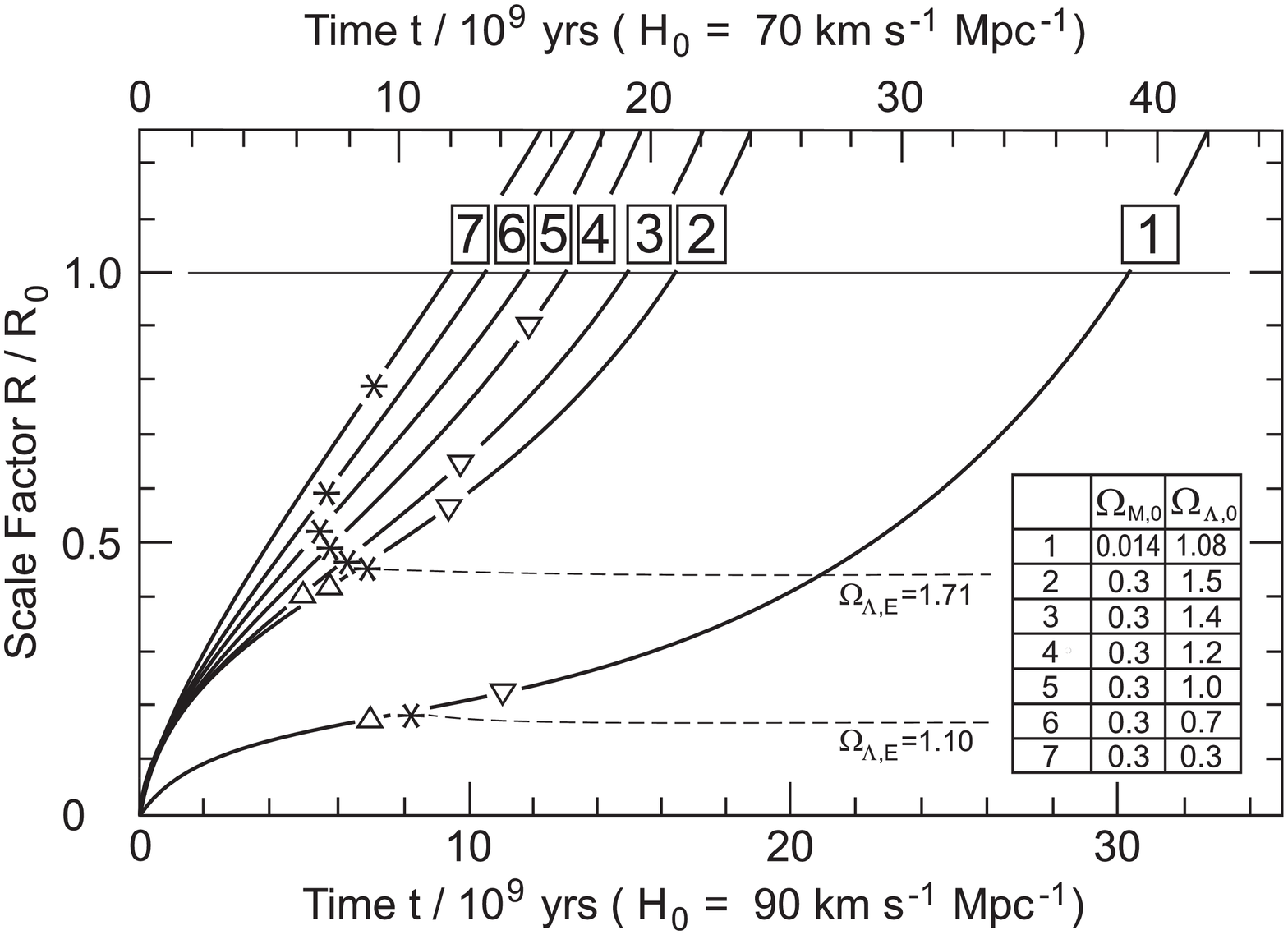}
\end{center}
\footnotesize{Fig.~2.  Evolution of the cosmological scale factor 
   $R(t)/R_0$ as a function of time in several models with $\Lambda > 0$}
\end{figure}

Several examples are plotted in Fig.~2, including closed models (1 through 5)
and one flat (6) and open model (7).  Model~1, with $\Omato=0.014$ and 
$\Olamo=1.08$, has been proposed in \cite{Lie92a,Lie92b} and will be 
discussed further in \S\ref{sec:bnp}.
The others all have $\Omato=0.3$, a figure widely
quoted today for the total density of gravitating matter (\S\ref{sec:exo}; 
see, e.g., \cite{Bah99}).  Model~6, with $\Olamo=0.7$ (known as the 
$\Lambda$CDM model), has been singled out as the newest ``standard model''
of cosmology.  Two timescales are plotted (top and bottom),
depending on the present value $\Hub$ of Hubble's parameter
(\S\ref{sec:bar}).

Along each of the curves, we have marked the points where
$\Omat(z)$ takes on maximum values ($\vartriangle$), the points of 
inflection ({\large $\ast$}), and the points where $\Olam(z)$ takes 
on maximum values and $H(z)$ reaches a minimum ($\triangledown$).
One finds that $R$ takes the special value 
$\bs{R}{\Lambda}\equiv 1/\sqrt{\Lambda}$ at the points ($\vartriangle$).
The cosmological constant may thus be understood physically
(in closed models) as the curvature of space at the time 
when the matter density parameter goes through its maximum 
(see \cite{Pri95} for discussion).

Joining the points of inflection in Fig.~2 are two dashed lines 
marked $\Oein$ (for ``Einstein limit'').  One must have $\Olamo < \Oein$ 
(a function of the matter density $\Omato$) in order for expansion to 
originate in a big bang singularity.
Solutions with $\Olamo=\Oein$ go over to Einstein's static model as 
$t\rightarrow\infty$.  When $\Olamo > \Oein$, $R(t)$ drops to a 
nonzero minimum and starts to climb again in the past direction; 
these are usually known as Eddington-Lema\^{\i}tre (or ``bounce'') models.

The value of $\Oein$ can be computed for a given model by differentiating
Eq.~(\ref{FE}) at the points of inflection ($\ast$).  This leads to a 
cubic polynomial which must be solved parametrically in general 
\cite{Fel86,Pri98} but has a little-appreciated direct solution for 
cases in which $\Omato\leqslant 0.5$ \cite{Blo85}.  With the conviction 
that $\Omato=1$ now fading in the astronomical community, and most 
cosmologists calling for values of $\Omato\approx 0.3$, it may be worthwhile 
to dust off this formula again.  In modern form it reads \cite{Blo97,Whi96}
\beqa
\Oein \!\!\! & = & \!\!\! 1 - \Omato + {\textstyle \frac{3}{2}} \, 
   \Omato^{\, 2/3} \nonumber \\
   & & \!\!\! \times \left[ \left( 1 - \Omato + \sqrt{1-2\Omato} \right)^{1/3}
   \right. \nonumber \\
   & & \!\!\! + \left. \left( 1 - \Omato - \sqrt{1-2\Omato} \right)^{1/3} 
   \right] \; .
\eeqa
One finds that $\Oein=1.10$ if $\Omato=0.014$, for instance 
(\S\ref{sec:bnp}), while $\Omato=0.06$ leads to $\Oein=1.25$.
Combined analysis of data on cosmic microwave background (CMB) fluctuations
from the {\sc Cobe}, {\sc Boomerang} and {\sc Maxima} experiments implies that
$\Ototo\leqslant 1.24$ at 2$\sigma$ confidence \cite{Jaf00}.  Since $\Olamo$
is certainly less than $\Ototo$, we infer that $\Olamo < \Oein$ in models with
$\Omato\geqslant0.06$, which may be regarded as a proof of the existence of 
the big bang in these models \cite{Ehl89}.  For higher matter densities 
$0.2 \leqslant\Omato\leqslant 0.5$ (as discussed in \S\ref{sec:exo}),
one obtains even larger values of $\Oein$, between 1.5 and 2.0
(see the Einstein limits in \cite{Pri98}).

The differences between the models shown in Fig.~2 become apparent
when their evolution is plotted on a phase diagram, with matter density 
parameter along one axis and vacuum density parameter along the 
other.  The key equations \cite{Blo97} are
\beqa
\Olam(z) \!\!\! & = & \!\!\! \Olamo / \left[ \Omato(1+z)^3 + 
   \Olamo \right. \nonumber \\
   & & \!\!\! -\left. (\Omato + \Olamo - 1)(1+z)^2 \right] \nonumber \\
\Omat(z) \!\!\! & = & \!\!\! (\Omato / \Olamo) \, \Olam(z) \, (1+z)^3 \; .
\eeqa
Fig.~3 depicts the same family of models as Fig.~2, with
redshift factors [$1+z=\bs{R}{0}/R(t)$] labelled at intervals along the curves.
Also marked are contours of constant {\em deceleration\/}, defined by 
$q\equiv -\ddot{R}R/\dot{R}^2 = \Omat/2-\Olam$.  This parameter takes 
values of 0.5 at each point of maximum matter density parameter 
($\vartriangle$), zero at the inflection points ($\ast$), and 
$-1$ at the points of minimum expansion rate ($\triangledown$).

\begin{figure}[t!]
\begin{center}
\includegraphics[width=85mm]{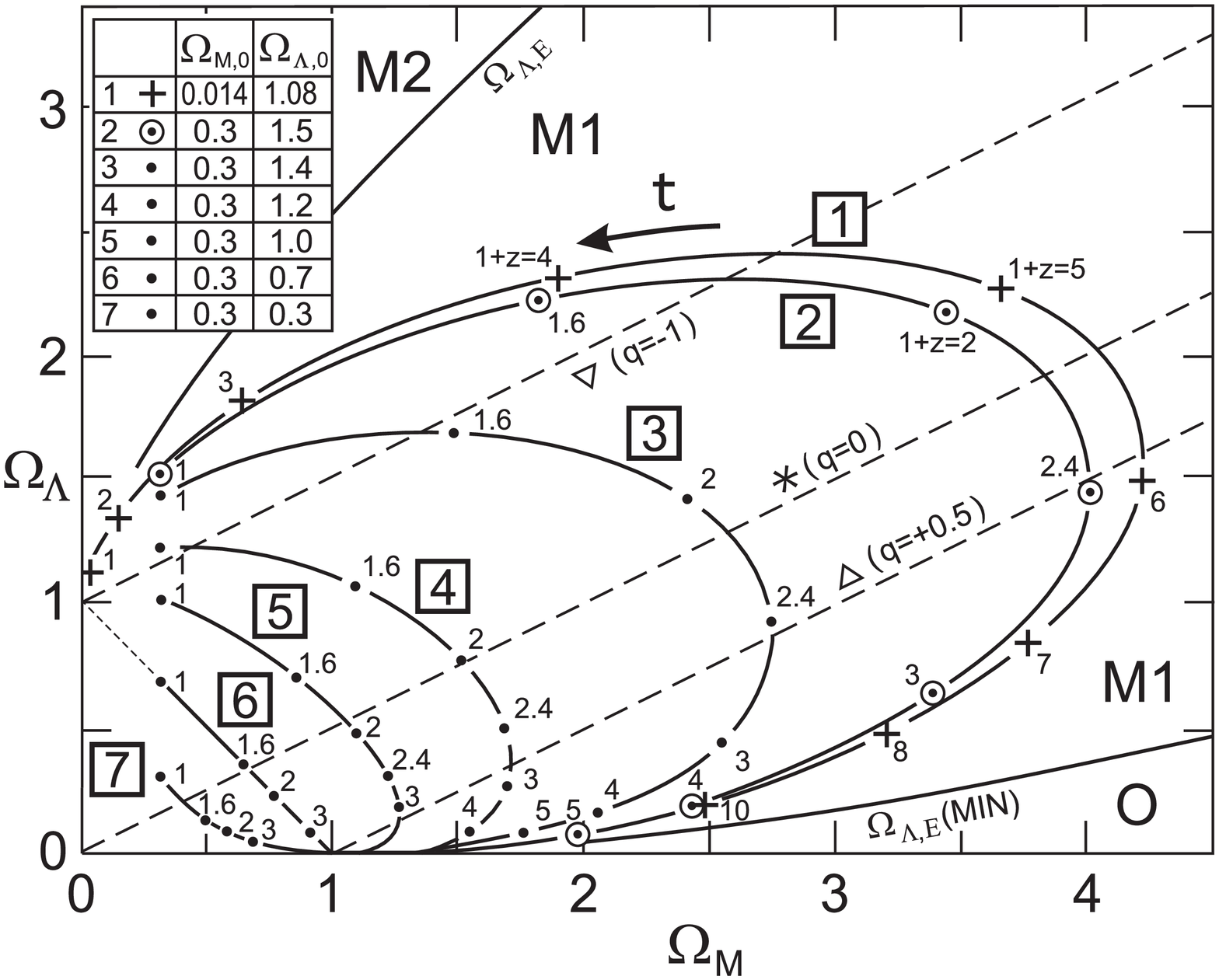}
\end{center}
\footnotesize{Fig.~3.  Evolution in phase space defined by 
   $\Omat(z)$ and $\Olam(z)$ of the models whose scale factors are 
   plotted in Fig.~2.  Plots of this kind go back at least to \cite{Sta66},
   and have been extended into the $\bs{\Omega}{R}$-direction in 
   \cite{Ehl89}.}
\end{figure}

All positive-$\Lambda$ models begin in Fig.~3 at the point (1,0)
and evolve asymptotically toward (0,1) as $t\rightarrow\infty$.
Flat (Euclidean) models follow a straight line;
any deviation from critical density produces a curved path.
Those to the right of Model~6 are all closed.  Models~5 through 2
are increasingly unlikely insofar as they violate the above-mentioned 
observational bound $\Ototo\leqslant 1.24$ on total density \cite{Jaf00}.
Model~2, in particular, cannot describe the real universe.
However, the model immediately adjacent to it in phase space (Model~1)
is perfectly acceptable in this regard, since it has $\Ototo=1.094$.
Very different combinations of $\Omato$ and $\Olamo$, in other
words, can produce almost identical trajectories in phase space.
Indeed, from the perspective of Fig.~3, the popular Euclidean models
appear implausibly fine-tuned.  One would not expect the Universe 
to take the shortest path through phase space, any more than one 
would expect a star to follow a straight line from the main sequence 
to the red giant branch on a Hertzsprung-Russell 
diagram.\footnote{Of course, the Universe we have assumed (homogeneous and
   isotropic) is simpler than most stars, and might be restricted to such 
   a path for reasons having to do with some higher symmetry of nature.
   While this may be so, however, any such symmetry would lie outside the
   context of Einstein's theory of gravity as it stands.}

The slow expansion rate and high matter density parameter between the points 
marked ($\vartriangle$) and ($\triangledown$) single out this stage of 
evolution for {\em large-scale structure formation\/}.  If $\Olamo$ 
is of the same order (or less) than $\Omato$, however, this process
must occur very quickly.  Consider $\Lambda$CDM, represented by Model~6
in Fig.~2.  Analysis of the Hubble and William Herschel Deep Fields
suggests that the number density of galaxies at redshifts
$z\approx 4-6$ (i.e., at scale factors $R/\bs{R}{0}\approx 0.1-0.2$)
is equal to that at $z=0$ \cite{Sha01}.  If so, then these objects
were in place by $z=4$, or (consulting Fig.~2 for Model~6, and using 
either the top or bottom scale for $\Hub$) {\em within 1.2-1.5~Gyr
after the big bang\/}.  This poses a serious challenge for the model,
since primordial density fluctuations must not only decouple
from the Hubble expansion on short timescales, but do so at a time 
when the expansion rate (the slope of the curve in Fig.~2) is some 
six times its present value.  The problem is even worse in models
with lower values of $\Olamo$ (e.g., Model~7).

The standard way to address this has been to suppose
that most of the matter density is in an exotic new form (CDM)
which is able to decouple from the primordial fireball before the baryons,
preparing potential wells for them to fall into (\S\ref{sec:exo}).  
This approach successfully accelerates structure formation on large 
scales \cite{Pee93}, but is perhaps {\em too\/} successful on smaller
ones.  Galaxy-sized regions are formed with excessively peaked 
central masses (the ``density cusp problem'') \cite{Moo94} and
too many low-mass fragments (the ``substructure problem'') \cite{Kly99}.
These problems may be resolved within the CDM picture by refining the 
properties of the exotic matter; proposals include warm \cite{Hog00}, 
fuzzy \cite{Hu00}, fluid \cite{Pee00a} and
self-interacting dark matter \cite{Spe00}.

Alternatively, difficulties with the growth of large-scale structure
are substantially eased in models with
larger ratios of $\Olamo$ to $\Omato$ \cite{Fel93,Sah92}.  In Model~1, for
instance, Fig.~2 shows that redshift $z=4$ corresponds to between
9.3 and 11.9~Gyr after the big bang (depending on the value of $\Hub$),
giving the galaxies seven times longer to form.
The expansion rate at this redshift is only 0.7 times its present value.
Nor is the low (present) density of gravitating matter a problem in this
model, because $\Omat(z)$ reaches levels as high as four times the critical
density at redshifts near $z\approx 5$ (Fig.~3).
It is natural to associate this redshift with the onset of galaxy formation,
and it would be of great interest to test ``slightly-closed'' cosmologies
of this kind via numerical simulations.

\begin{figure}[t!]
\begin{center}
\rotatebox{270}{\includegraphics[width=105mm]{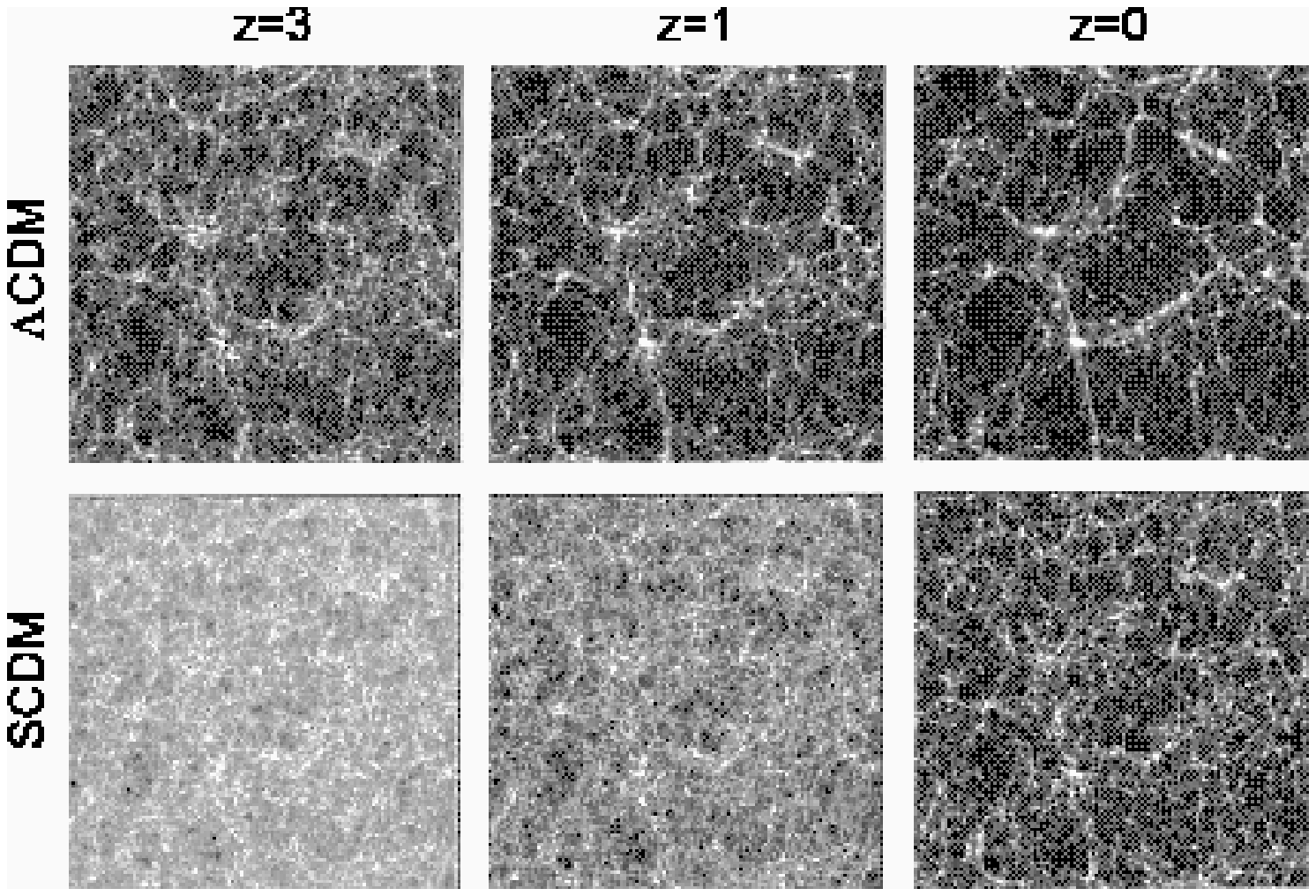}}
\end{center}
\footnotesize{Fig.~4.  Numerical simulations of structure formation.
   On the left is the ``old standard model'' (SCDM) with $\Omato=1,
   \Olamo=0$ and $\hub=0.5$.  On the right is the ``new standard model''
   ($\Lambda$CDM) with $\Omato=0.3, \Olamo=0.7$, and $\hub=0.7$.
   Panel size is scaled to match the Hubble expansion; time runs from 
   top ($z=3$) to bottom ($z=0$).  (Images courtesy J.~Colberg and the
   {\sc Virgo} Consortium).}
\end{figure}

Existing studies like that shown in Fig.~4 by the {\sc Virgo} Consortium
\cite{Jen98} have so far been restricted to flat and open models.
Of these, $\Lambda$CDM (the new ``standard CDM'' model) produces a 
mass distribution in much better agreement with the observed distribution 
of light than EdS (the old ``standard CDM'' model, still denoted SCDM 
in many works).  The improvement is especially pronounced at higher 
redshifts (top two panels).  This has been taken as evidence for a
significant $\Lambda$-term.  More detailed analysis, however, reveals
that the power spectrum of the mass distribution does not agree with
that observed for galaxies in {\em either\/} of these two models
\cite{Jen98}.  The discrepancies have typically been attributed to
nonlinear, scale- and morphology-dependent bias factors \cite{Bla99}; 
a complementary approach would be to consider an expanded repertoire 
of $\Omato$ and $\Olamo$ values.

\section{Baryons} \label{sec:bar}

Beginning from the observed luminosity density of the Universe, one can
infer the total density of {\em luminous baryonic matter\/} (i.e., that
in stars) by making various reasonable assumptions about the fraction of
galaxies of different morphological type, their ratios of disk-type
to bulge-type stars, and so on.  The latest such estimate is \cite{Fuk98a}
\beq
\Olum = (0.0027 \pm 0.0014) \hub^{-1} \; ,
\label{raw-Olum}
\eeq
where $\hub$ is the present value of Hubble's parameter expressed
in units of 100~km~s$^{-1}$~Mpc$^{-1}$.

This latter parameter is unfortunately not yet fixed by observation,
and we pause to discuss its value before proceeding.  Using various 
relative-distance methods, all calibrated against Cepheid variable stars 
in the Large Magellanic Cloud (LMC), the Hubble Key Project ({\sc Hkp}) team
has determined that $\hub = 0.71 \pm 0.06$ \cite{Mou00}.  Fundamental
physics methods (e.g., the Sunyaev-Zeldovich effect, gravitational 
lensing time delays [GLTDs]) have higher uncertainties but are roughly
consistent with this, $\hub = 0.65 \pm 0.08$ \cite{Pri00}. 
The near convergence of these approaches has been widely hailed, with many
authors asserting that precision values of $\hub$ are just around the corner,
awaiting only a final round of experimental refinements.

On the other hand, a recalibrated LMC Cepheid period-luminosity relation
based on a much larger Cepheid sample (from the {\sc Ogle} collaboration)
leads to a considerably higher value of $\hub = 0.85 \pm 0.05$ \cite{Wil00a}.
And a new VLBI measurement of the transverse velocity of water masers in
NGC4258 gives a purely geometric distance to this galaxy \cite{Her99}
which also implies that the traditional calibration is off, boosting
Cepheid-based estimates by $12 \pm 9$\% \cite{Mao99}.  This would raise,
e.g., the {\sc Hkp} value to $\hub = 0.80 \pm 0.09$.  There is some
additional independent support for this recalibration in observations of
eclipsing binaries \cite{Gui98} and ``red clump stars'' in the LMC
\cite{Uda00}.  Fundamental physics approaches are also not immune to
systematic effects: GLTD-based values of $\hub$, which are routinely
computed assuming EdS, rise by about 7\% (on average) in $\Lambda$CDM, 
and 9\% in open models.

``Hubble fatigue'' may therefore have prompted cosmologists to embrace 
prematurely small levels of uncertainty in $\hub$.  We attempt to allow
for this by retaining two possible values for $\hub$ in this review;
a ``low value'' of $\hub=0.7$ and a ``high value'' of $\hub=0.9$.
This seemingly modest range of choices turns out to discriminate quite
powerfully between the cosmological models considered here.  To a large
extent this is a function of their ages.  Fig.~2 reveals, for example,
that $\Lambda$CDM (represented by Model~6) is 13.5~Gyr old if $\hub=0.7$,
or 10.5~Gyr if $\hub=0.9$.  The oldest metal-poor halo stars seen in the
Milky Way have an age of $15.6 \pm 4.6$~Gyr \cite{Cow99}, setting a firm
lower limit of 11.0~Gyr on the age of the Universe.  This is enough to
rule out $\Lambda$CDM with the high value of $\hub$, but not the low one.

Model~1, on the other hand, faces the opposite problem: Fig.~2 shows
that it has a total age of 30.2~Gyr if $\hub=0.9$, or 38.8~Gyr if $\hub=0.7$.
Both numbers are larger than most cosmologists are prepared to accept.
However, upper limits on the age of the Universe are not as secure
as lower ones.  One must take into account, for instance, that the
galaxy formation associated above with redshifts $z\approx 4$
occured between 9 and 12~Gyr after the big bang in this model,
so that galaxies would not be older than $24 \pm 3$~Gyr in any case.

There are various ways to test such a hypothesis.  One might expect to
find a greater {\em spread\/} in galaxy ages (and hence colors) at 
$z \approx 4$, given their longer formation time.  Galaxies in Model~1
had $\sim$6~Gyr to form, or about $\sim$25\% of their nominal lifetime, 
according to Fig.~2 (with $\hub=0.9$).
The corresponding fraction in Model~6 is $\lesssim$10\%.
This may not necessarily translate into a difference between 
observed color spreads, however, since high-redshift 
galaxies are seen principally during (relatively brief) episodes 
of star formation.

Very old galaxies, if they exist, should also be present at lower redshifts.
They would be inherently faint and reddened, making them difficult
to find and distinguish from younger objects which are simply obscured 
by dust.  Nevertheless, several candidates have been noted in the past 
years, including a number of low surface brightness galaxies \cite{Ber00} 
and extremely red objects \cite{Coh99,Hu94} whose ages based on simple 
(``single-burst'') evolution models appear to be as high as $\sim$17~Gyr 
or more.

If our own Milky Way is not unusually young, we should also expect to find
large numbers of dead stars in the galactic halo.  These would act as 
{\em microlenses\/}, inducing variability in background stars and 
quasars, even if they were too dim to be seen directly.  Such objects may
now have been detected in the direction of the LMC (see below).

Returning now to the density of luminous matter, we find with our
values for $\hub$ that Eq.~(\ref{raw-Olum}) gives
\beq
\Olum = 0.0034 \pm 0.0018 \; .
\label{Olum}
\eeq
The visible Universe, in other words, constitutes an insignificant 0.5\%
or less of the critical density.

It may, however, be that many of the baryons in the Universe are not
visible.  How significant could these dark baryons be?
The theory of cosmic nucleosynthesis provides us with an independent
method for determining the density of {\em total\/} baryonic matter 
in the Universe, based on the assumption that the light elements we see
today were forged in the furnace of the hot big bang.
Results using different light elements agree tolerably well, which is
impressive in itself.  The primordial abundances of $^4$He
(by mass) and $^7$Li (relative to H) imply a baryon density of 
$\Obar = (0.011 \pm 0.005) \hub^{-2}$ \cite{Oli00}, whereas
new measurements based exclusively on the primordial D/H abundance
give a higher value: $\Obar = (0.019 \pm 0.002) \hub^{-2}$ \cite{Tyt00}.
These two results, both given at the 2$\sigma$ confidence level, are 
superimposed on a plot of predicted light element abundances in Fig.~5.
\begin{figure}[t!]
\begin{center}
\includegraphics[width=70mm]{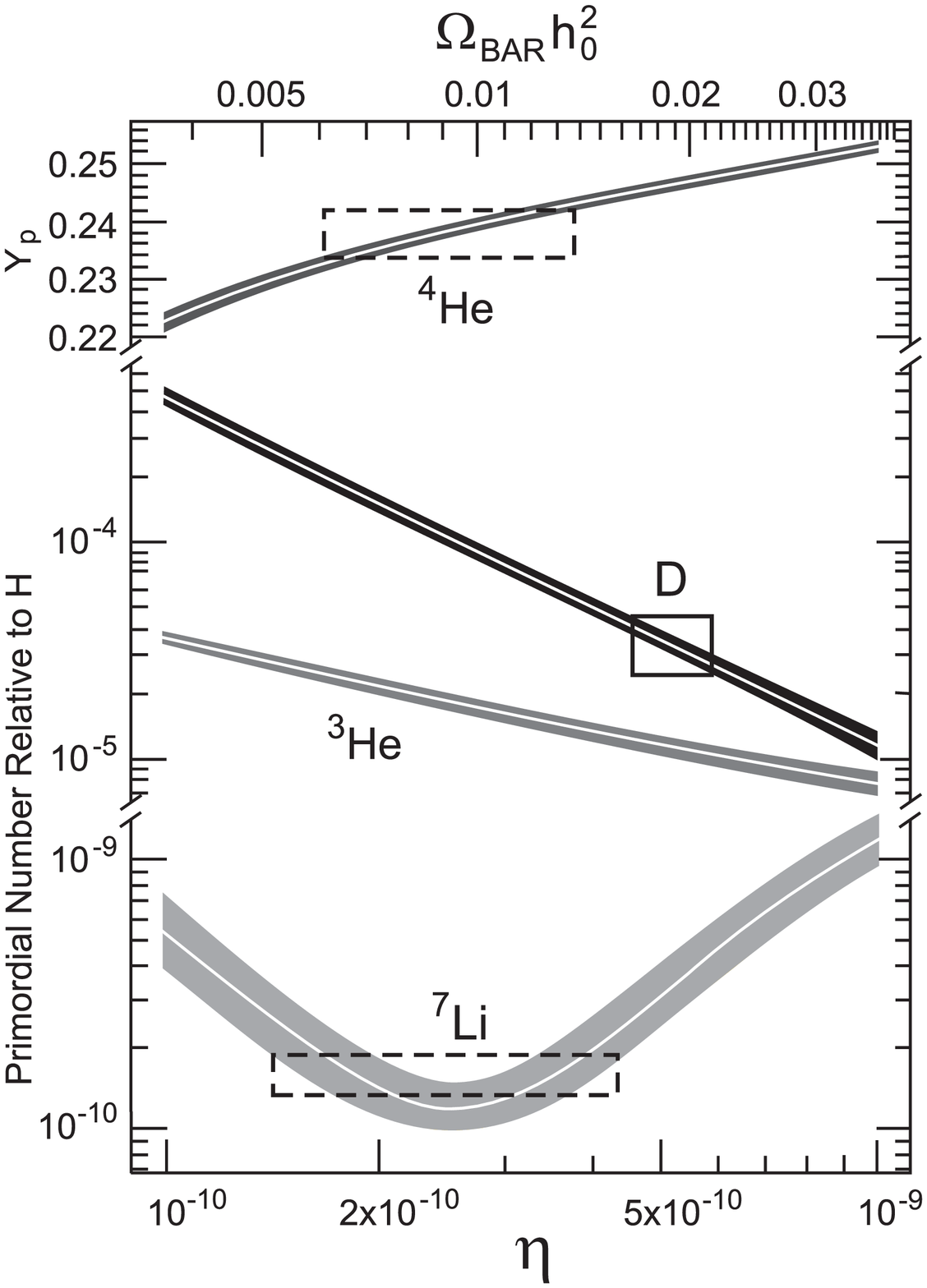}
\end{center}
\footnotesize{Fig.~5.  Measurements of the primordial abundance
   of light elements $^4$He, D and $^7$Li (scales on left-hand side) determine
   the value of $\Obar$ (top).  The bottom scale gives the corresponding 
   baryon-to-photon ratio $\eta$ (adapted from \cite{Oli00,Tyt00}).}
\end{figure}
They do not overlap.  In our view it is premature at present to exclude 
either one.  We therefore adopt the same strategy here as with Hubble's 
parameter, retaining a ``low'' baryon density of $0.01 \hub^{-2}$ and 
a ``high'' one of $0.02 \hub^{-2}$ throughout our review.  Combining 
this with our high and low values of $\hub$, we conclude that the 
baryonic density lies in the range
\beq
\Obar = 0.012 - 0.041 \; ,
\label{Obar}
\eeq
a result in very good agreement with that obtained by the entirely
independent method of adding up individual contributions from all
known repositories of baryonic matter via their estimated
mass-to-light $(M/L)$ ratios \cite{Fuk98a}.  If $\Ototo$ is close to
unity, as it now seems (\S\ref{sec:vac}), then it follows from
Eq.~(\ref{Obar}) that baryons --- and everything that would have
been recognized as ``matter'' before 1930 --- make up less than
5\% (by mass) of the known Universe.

Most of these baryons, moreover, have not been seen.  The baryonic
{\em dark\/} matter fraction $\bs{f}{BDM} (\equiv\Obdm/\Obar) = 
1-\Olum/\Obar$ lies in the range
\beq
\bs{f}{BDM} = 77\% - 95\% \; ,
\eeq
where we have used Eq.~(\ref{raw-Olum}) together with our high and low
values of $\hub$ and $\Obar$.  Where could these dark baryons be?
One possibility is that they are smoothly distributed in a gaseous
intergalactic medium, which would have to be strongly ionized in order 
to explain why it has not left a more obvious signature in quasar absorption
spectra.  Recent observations using {\sc Ovi} absorption lines as a tracer
of ionization suggest that the contribution of such material to $\Obar$
is at least $0.003 \hub^{-1}$ \cite{Tri00}, comparable to $\Olum$.
Numerical simulations are able to reproduce many observed features 
of the Lyman $\alpha$ (Ly$\alpha$) forest with as many as $80-90$\% 
of the baryons in this form \cite{Mir96}.

Dark baryonic matter, however, could also be bound up in clumps
of matter such as substellar objects (jupiters, brown dwarfs) or
stellar remnants (white, red and black dwarfs, neutron stars, black holes).
The former are not thought to be numerous enough to be important, 
given their low mass.  The latter are limited in the opposite sense;
black holes cannot be more massive than about $10^5 M_{\odot}$ since this
would lead to dramatic tidal disruptions and lensing effects which are 
not seen \cite{Tur91}.  The critical mass range for dark baryon clumps 
is thus within a few orders of magnitude of the solar mass.  Microlensing
constraints based on quasar variability do not seriously limit such 
objects at present, setting an upper bound of 0.1 (well above $\Obar$)
on their combined contributions to the density parameter of an EdS 
universe \cite{Sch93}.

A likely detection of dark compact objects within our own galactic halo
has recently been reported by the {\sc Macho} microlensing survey of 
stars in the LMC \cite{Alc00}.
The lenses, with masses in the range $(0.15-0.9)M_{\odot}$, appear to
account for between 8\% and 50\% of the high rotation velocities seen in 
the outer parts of the Milky Way --- depending on the halo model chosen,
and extrapolating (at 2$\sigma$ confidence) from the $\sim$15 events actually 
seen.\footnote{An alternative interpretation of the data, that most or
   all of these lenses are actually faint stars inside the LMC itself,
   now appears to be strongly disfavored \cite{Nel00}.}
The identity of these objects has been hotly debated,
with some authors linking them to faint, fast-moving objects 
apparently detected in the Hubble Deep Field \cite{Iba99}.  It is unlikely
that they could be traditional white dwarfs, since these are formed from
massive progenitors whose metal-rich ejecta we do not see \cite{Fie00}.
Degenerate ``beige dwarfs,'' which might be able to form above the
hydrogen-burning mass limit of 0.08 $M_{\odot}$ without fusing \cite{Han99}
are one possibility.  Another would be a population of ancient, low-mass
($\lesssim 0.6 M_{\odot}$) stars which have simply cooled into
invisibility.\footnote{Existing limits on the density of halo objects 
   in this mass range \cite{Gra96} necessarily extrapolate in a number of
   ways from known stellar populations, and are based on theoretical stellar
   evolution models \cite{Ale97,Cha99} which do not currently extend beyond
   20~Gyr.}

\section{Exotic Dark Matter} \label{sec:exo}

Three main reasons have been proposed for going beyond dark baryons and 
introducing a second species of invisible matter, the 
{\em exotic cold dark matter\/} (CDM), into the Universe: 
(1) a range of observational arguments imply that the density parameter
of total gravitating matter ($\Omato=\Obar+\Ocdm+\Onu$) is higher than
that provided by baryons and neutrinos alone; 
(2) our current understanding of large-scale structure formation requires
the process to be helped along by quantities of {\em cold\/} (i.e., 
nonrelativistic) gravitating matter in the early universe, creating the
potential wells for infalling baryons; and
(3) theoretical physics supplies several plausible (albeit still hypothetical)
candidate CDM particles with the right properties.

Since our ideas on structure formation may yet change,
and the candidate particles may not materialize, the case for exotic CDM
turns on the observational arguments. 
At present these agree to within no better than a factor of five, 
pointing to values of $\Omato$ between about 0.1 and 0.5.
(Not long ago, in the 1980s, there were calls for $\Omato=1$,
but these tended to come from theorists wishing to retain the
EdS model, and are no longer tenable observationally \cite{Whi93}.)
The lower limit is crucial: if $\Omato > \Obar+\Onu$, then $\Ocdm > 0$.

The arguments can be broken into two classes: those which are purely 
empirical, and those which assume in addition the validity of the 
gravitational instability (GI) picture of structure formation.
Let us begin with the empirical arguments.  One has been encountered
already: the {\em galactic rotation curve\/}.  If the {\sc Macho} results
are taken at face value, and if the Milky Way is typical, then it is probable
that dark compact objects make up less than 50\% of the halo mass in 
spiral galaxies.  The remaining halo dark matter does not appear to 
consist of baryonic matter in known forms such as dust, rocks, hot 
or cold gas, or hydrogen snowballs \cite{Heg86}.

The total amount of dark matter in spiral galaxies, however, is rather
limited.  The easiest way to see this is to divide the total dynamical 
mass of the Milky Way (including its unseen halo matter) by its 
luminosity.\footnote{$\bs{M}{MW}$ has been calculated at
   $(4.9 \pm 1.1) \times 10^{11} M_{\odot}$ inside 50~kpc using
   using the motions of galactic satellites \cite{Koc96a}; while
   $\bs{L}{MW}$ is given by $(2.3 \pm 0.6) \times 10^{10} L_{\odot}$ 
   in the $B$-band.}
The resulting mass-to-light ratio, $M/L=(21 \pm 7) \MLunits$,
is a mere $\sim0.02\hub^{-1}$ times that of a critical-density universe, 
$\bs{(M/L)}{crit} = (1136 \pm 138)\hub\MLunits$ \cite{Car97}.

Most of the mass of the Universe, in other words, is spread over scales 
larger than galaxies, and it is here that the arguments for exotic CDM
are most compelling.  The {\em $M/L$-ratio\/} method is in fact the 
most straightforward:  one measures $M/L$ for a chosen region, corrects
for the corresponding value in the ``field,'' and divides by 
$\bs{(M/L)}{crit}$ to obtain $\Omato$.  Much, however, depends on 
the region.  A widely respected application of this approach, that 
of the {\sc Cnoc} team, uses rich clusters of galaxies.
These systems sample large volumes of the early Universe, have dynamical
masses which can be measured by three independent methods
(the virial theorem, $x$-ray gas temperatures and gravitational lensing),
and are subject to fairly well-understood evolutionary effects.
They are found to have field $M/L$-ratios of $(213 \pm 59)\hub\MLunits$,
giving $\Omato = 0.19 \pm 0.06$ (1$\sigma$ confidence) when 
$\Olamo = 0$ \cite{Car97}.  This result scales as $(1-0.4\Olamo)$ 
\cite{Car99} so that, e.g., $\Omato$ drops to $0.11 \pm 0.04$ 
when $\Olamo = 1$.

The weak link in this chain of inference is that rich clusters may not
be characteristic of the Universe as a whole.  Only about 10\% of galaxies
are found in such clusters.  If {\em individual\/} galaxies 
(like the Milky Way, with $M/L \approx 21$) are substituted for clusters,
then the inferred value of $\Omato$ drops by a factor of ten, 
approaching $\Obar$ (\S\ref{sec:bar}) and removing the need for exotic CDM.
A recent comprehensive effort to address the impact of scale on $M/L$ 
arguments concludes that $\Omato = 0.16 \pm 0.05$ when regions of 
{\em all\/} scales (from individual galaxies to superclusters)
are considered \cite{Bah00}.\footnote{This result is based on comparison
   with simulations that assume $\Omato=0.37$ plus spatial flatness.
   It is hence subject to systematic uncertainty (in the downward direction) 
   of as much as -20\% \cite{Bah00}, possibly more if $\Olamo > 0.63$.}

A second line of argument uses the {\em cluster baryon fraction\/}
($\barf$) of baryonic to total gravitating mass in clusters.  Baryonic 
matter is defined as the sum of visible galaxies and hot cluster gas
(the mass of which can be inferred from its $x$-ray temperature).
Total cluster mass is measured by one or all of the three methods 
listed above (virial, $x$-ray, or lensing).  At sufficiently large radii
the cluster may be taken as representative of the Universe as a whole, 
so that $\Omato = \Obar/(\barf)$, where $\Obar$ is fixed by
nucleosynthesis (\S\ref{sec:bar}).  Applied to various clusters, this
procedure leads to $\Omato = 0.3 \pm 0.1$ \cite{Bah99} ---
a result which is almost certainly an upper limit, partly because
baryon enrichment is more likely to take place inside the cluster than out,
and partly because {\em dark\/} baryonic matter (e.g., {\sc Macho}s) is not 
taken into account; this would raise $\bs{M}{BAR}$ and lower $\Omato$.

A final, recent entry into the list of purely empirical methods
uses the separation of {\em radio galaxy lobes\/} as standard rulers,
a variation on the classical angular-size distance test in cosmology. 
The widths, propagation velocities and magnetic field strengths 
of these lobes have been measured for 14 radio galaxies with 
the aid of long-baseline radio interferometry, leading to the (1$\sigma$)
constraint $\Omato < 0.10$ for $\Olamo=0$, or 
$\Omato = 0.10^{+0.25}_{-0.10}$ for flat models ($\Olamo = 1 - \Omato$) 
\cite{Gue00}.

We consider next the GI-based measurements of $\Omato$, which are
``circular'' in the sense that they assume that large-scale structure formed
via gravitational instability from a Gaussian spectrum of primordial density
fluctuations --- a process which (as we currently understand it) could not
have taken place as it did {\em unless\/} $\Omato$ is considerably larger
than $\Obar$.  According to GI theory, formation of large-scale structure
is more or less complete by $z \approx \Omato^{-1} - 2$ \cite{Pad93}.
Therefore one can constrain $\Omato$ by looking for evidence of
{\em number density evolution\/} of large-scale structures such as
galaxy clusters.  In a low-density universe, this would be relatively
constant out to at least $z\sim1$, whereas in a high-density universe
one would expect the abundance of clusters to drop rapidly with $z$ because
they are still in the process of forming.  In fact, massive clusters are
seen at redshifts as high as $z=0.83$, leading to the (1$\sigma$) limits 
$\Omato = 0.17^{+0.14}_{-0.09}$ for $\Olamo=0$ models, and
$\Omato = 0.22^{+0.13}_{-0.07}$ for flat ones \cite{Bah98}.

Evolution of the {\em mass power spectrum\/} $P(k)$ constrains $\Omato$ 
in a similar way.  Here one uses the fact that structures of a given mass
form by the collapse of large-scale regions in a low-density universe,
or smaller-scale regions in a high-density one.
Comparing $P(k)$ for the present-day distribution of galaxy clusters to that
for the distribution of matter at some earlier epoch therefore yields 
an estimate of $\Omato$.  Using the mass power spectrum of Ly$\alpha$ 
absorbers at $z\approx2.5$, for instance, one finds that
$\Omato = 0.46^{+0.12}_{-0.10}$ (1$\sigma$) for $\Olamo =0$ models
\cite{Wei99}.  This result goes as approximately $(1-0.4 \Olamo)$,
so that the central value of $\Omato$ drops to 0.34 in a flat model, 
and 0.28 if $\Olamo=1$.

A final group of measurements comes from galaxy {\em peculiar velocities\/}.
These are produced by the gravitational potential of locally over-
(or under-) dense regions relative to $\Omato$, but also depend on $\Omato$
itself.  The power spectra of the velocity and density distributions 
can be related within the context of GI theory.  A typical bound derived from
several such studies is $\Omato > 0.3$ \cite{Zeh99}.  Dependence on $\Olamo$
is modest since these tests probe relatively small volumes, but lower limits
derived in this way can depend significantly on $\hub$ as well as the spectral
index $n$ of the density distribution.  In \cite{Zeh99}, where the latter
is normalized to CMB fluctuations, results take the form
$\Omato \hub^{1.3} n^2 \approx 0.33\pm0.07$ (2$\sigma$).
The preferred value of $\Omato$ therefore drops from $0.53 \pm 0.11$
(if $\hub=0.7$) to $0.38 \pm 0.08$ (if $\hub=0.9$),
where we have assumed $n=1$.

To summarize, one may say that purely empirical arguments lean toward
values of $\Omato\sim0.3$ or {\em lower\/}, whereas GI (gravitational
instability) theory-based results tend to come in at $\sim0.2$ and 
{\em higher\/}.  If there is flexibility in the lower limits on $\Omato$,
it lies in the empirical methods, especially if contributions from dark
baryons are near their upper limit (\S\ref{sec:bar}).
It is unlikely, however, that the limits based on GI theory can be
stretched far enough to remove the need for exotic CDM.  We therefore
conclude that this component of the dark matter has a density parameter
in the range
\beq
\Ocdm = \left\{ \begin{array}{cc}
   0.1 - 0.5 & \mbox{ (GI theory) } \\
   0 - 0.4   & \mbox{ (otherwise) } 
\end{array} \right.
\label{OCDM}
\eeq
If our current understanding of structure formation via gravitational
instability is correct, then exotic CDM must exist.
Conversely, if exotic CDM does not exist, then our understanding of 
structure formation is incomplete.

The debate, of course, becomes moot if exotic CDM (with $\Ocdm\sim0.3$)
is actually discovered in the laboratory.
Theorists have proposed a colorful list of particle candidates,
with varying degrees of testability.  Two have emerged as most
plausible: the axion and the weakly interacting massive particle 
(WIMP).\footnote{Other contenders include WIMPzillas, primordial
   black holes, magnetic monopoles, solitons (in various dimensions),
   cosmic string loops, shadow matter and mirror matter.  We refer the
   interested reader to the latest conference reports \cite{Kla99,Kla00} 
   or semipopular works such as \cite{Kra00,Rio91}.}
Either one of these could in principle make up the CDM because
each, if it exists, is (1) {\em cold\/} (i.e., nonrelativistic in the 
early Universe), and (2) {\em expected\/} on theoretical grounds to have
a collective density within a few orders of magnitude of the critical 
density (see {\bf Appendix}).
Ambitious experimental detection efforts around the world are now
directed at both particles.  While they have not turned up anything so far,
most of the theoretical parameter space remains unexplored.

\section{Neutrinos} \label{sec:nu}

Since neutrinos indisputably exist, and in great numbers (their number
density $n_{\nu}$ is $3/11$ times that of the CMB photons, or 112~cm$^{-3}$
per species), they have been leading particle dark matter candidates for
longer than either the axion or the WIMP.  They gained prominence in 1980
when teams in the U.S.A. and Soviet Union both claimed to have evidence
of nonzero neutrino rest masses.  While these claims did not stand up,
a new round of more sophisticated experiments is once again suggesting
that $m_{\nu}$ (and hence $\Onu) > 0$. 

Dividing $n_{\nu} m_{\nu}$ by the critical density of the Universe,
one obtains immediately \cite{Pee93}
\beq
\Onu = (\sum m_{\nu} c^2/94 \mbox{ eV}) \hub^{-2} \; ,
\label{Om_nu}
\eeq
where the sum is over the three neutrino species.\footnote{The calculations
   in this section are strictly valid only for $m_{\nu} c^2 \lesssim$~MeV.
   More massive neutrinos with $m_{\nu} c^2\sim$~GeV were once considered as
   CDM candidates but are no longer viable since LEP experiments rule out
   additional neutrino species with $m_{\nu} c^2<46$~GeV (i.e., half the
   $\bs{Z}{0}$-mass).}
Current laboratory upper bounds on neutrino rest masses are 3~eV 
($m_{\nu_e}c^2$), 0.19~MeV ($m_{\nu_{\mu}}c^2$) and 18~MeV 
($m_{\nu_{\tau}}c^2$), so it would appear feasible in principle 
for these particles to close the Universe.  In fact $m_{\nu_{\mu}}$ and 
$m_{\nu_{\tau}}$ are limited far more stringently by Eq.~(\ref{Om_nu}) 
than by laboratory bounds.

Perhaps the best-known theory along these lines in recent 
years is that of Sciama \cite{Sci93}, who postulated a population of 
29~eV $\tau$-neutrinos, decaying via
$\nu_{\tau} \longrightarrow \nu_{\mu} + \gamma$ into much lighter
$\mu$-neutrinos and 15~eV photons on timescales of
$\tau_{\nu} \lesssim 3 \times 10^{23}$~s.
Decay photons with these properties would solve a number
of astrophysical puzzles, such as the high degree of ionization in the 
interstellar medium and the large intergalactic flux of hydrogen-ionizing
photons.  The proposed neutrinos would moreover provide exactly the
critical density if $\hub=0.56$.  This model, however, has now been ruled
out by the absence of a strong 15~eV emission line in the
extragalactic background light \cite{Ove00}.

More generally, the strongest upper limits on $\Onu$ come from our
current understanding of {\em structure formation\/}.
Neutrinos are hot (i.e., relativistic at the time of decoupling from
the primordial fireball) and therefore able to stream freely out 
of density perturbations in the early Universe, erasing them
before they have a chance to grow.  Good agreement with observations 
of large-scale structure can be
achieved in models with $\Onu$ as high as 0.2, but only if $\Obar+\Ocdm=0.8$
and $\hub=0.5$ \cite{Gaw98}.  A more realistic upper limit follows from 
a statistical exploration of the entire parameter space and leads to the
conclusion that $m_{\nu}c^2 \lesssim (9.2$~eV)$\Ocdm$ ($2\sigma$) over
$0\leqslant\Ocdm\leqslant0.6$ for flat models \cite{Cro99}. 
Eq.~(\ref{Om_nu}) then implies that $\Onu<0.12\Ocdm$ (if $\hub=0.9$) or
$0.20\Ocdm$ (if $\hub=0.7$) --- a neutrino density below that attributed 
to exotic CDM (\S\ref{sec:exo}), but still well above that of the
baryons (\S\ref{sec:bar}).\footnote{These (possibly quite significant)
   values of $\Onu$ do not invalidate our earlier assumption (\S\ref{sec:cos})
   that the present density of radiationlike matter is negligible.
   While neutrinos are relativistic at decoupling, they lose energy and
   become {\em non\/}relativistic on timescales $\bs{t}{NR} \approx 
   190,000 (m_{\nu}c^2/$eV)$^{-2}$~yr \cite{Kol90} --- well before the present 
   epoch for neutrinos which are massive enough to be of interest.}

Unexpected new {\em lower\/} limits on $\Onu$ have come from atmospheric
(Super-Kamiokande \cite{Fuk98b}), solar ({\sc Sage} \cite{Abd99},
Homestake \cite{Cle98}, {\sc Gallex} \cite{Ham99}), and accelerator-based
({\sc Lsnd} \cite{Ath98}) neutrino experiments.  In each case it appears
that two neutrino species are oscillating into each other, a process
which can only take place if $m_{\nu}>0$.  The strongest evidence
comes from Super-Kamiokande, which has reported oscillations
between $\tau$- and $\mu$-neutrinos\footnote{A second
   possibility, that of oscillations between $\nu_{\mu}$ and a sterile
   {\em fourth\/} neutrino species ($\nu_s$), has now been excluded at
   the 3$\sigma$ confidence level \cite{Fuk00}.}
with $5 \times 10^{-4} \mbox{ eV}^2 < \Delta m_{\tau\mu}^2 c^2 < 6 \times 
10^{-3} \mbox{ eV}^2$ (2$\sigma$), where $\Delta m_{\tau\mu}^2 \equiv 
|m_{\nu_{\tau}}^2-m_{\nu_{\mu}}^2|$ \cite{Fuk98b}.
If neutrino masses are hierarchical, like the masses of other fermions,
then $m_{\nu_{\tau}} \gg m_{\nu_{\mu}}$ and $m_{\nu_{\tau}} c^2>$~0.02~eV.
In this case it follows from Eq.~(\ref{Om_nu}) that
$\Onu > 0.0003$ (if $\hub=0.9$) or 0.0005 (if $\hub=0.7$).
If, instead, neutrino masses are nearly degenerate, then $\Onu$ cannot
be determined from this result, but will in any case still lie below 
the upper bound imposed by structure formation above.  We conclude 
that the possible range of values for this parameter is
\beq
\Onu = 0.0003 - 0.2\Ocdm \; .
\label{Onu}
\eeq
The neutrino contribution to $\Ototo$ is anywhere from an order of
magnitude below that of the visible stars and galaxies (\S\ref{sec:bar})
up to as much as half that attributed to exotic CDM (\S\ref{sec:exo}).
If $\Ocdm$ is small (or zero), however, then $\Onu$ must lie near the
{\em lower\/}, rather than the upper end of this range; since a large 
density of neutrinos will interfere with structure formation (as noted above) 
unless something like exotic CDM is present to help hold primordial
density perturbations together.

\section{Vacuum Energy} \label{sec:vac}

There are at least four good arguments for the cosmological constant.
The first is {\em mathematical\/}: $\Lambda$ plays a role in 
Eqs.~(\ref{EFEs}) similar to that of the additive constant in an
indefinite integral \cite{Rin77}.
The second is {\em dimensional\/}: $\Lambda$ specifies the curvature radius
$\bs{R}{\Lambda}\equiv 1/\sqrt{\Lambda}$ of a (closed) Universe at the 
moment when the matter density parameter $\Omat(z)$ passes through its 
maximum (\S\ref{sec:cos}), thereby providing a fundamental length scale
for cosmology (see \cite{Pri95} for discussion).
The third is {\em dynamical\/}: $\Lambda$ determines the asymptotic 
expansion rate of the Universe according to Eq.~(\ref{magic}), 
$\Lambda c^2 = 3 H_{\infty}^2$.
And the fourth is {\em material\/}:  $\Lambda$ is related to the energy
density of the vacuum via $\rlam c^2=\Lambda c^4/8\pi G$.

With all these reasons to take this term seriously, why have most 
cosmologists since Einstein set $\Lambda=0$?  Computational convenience 
is one explanation.  Another is the smallness of most effects associated
with the $\Lambda$-term.  Einstein himself set $\Lambda=0$ in 1931
``aus Gr\"unden der logischen \"Okonomie'' --- for reasons of 
logical economy --- because he saw no hope of measuring this 
quantity experimentally at the time.  He is often quoted as adding 
that its introduction in 1915 was the ``biggest blunder'' of his life
(``die gr\"o{\ss}te Eselei in meinem Leben'').
This statement, which does not appear anywhere in Einstein's writings
but was rather attributed to him by Gamow \cite{Gam70}, is sometimes
interpreted as a rejection of the very idea of a cosmological constant.
It more likely represents Einstein's rueful recognition that, by 
invoking the $\Lambda$-term solely to obtain a static solution of 
the field equations, he had narrowly missed what would surely have 
been one of the greatest {\em triumphs\/} of his life:
the prediction of cosmic expansion.

The relation between $\Lambda$ and the energy density
of the vacuum has led to a new quandary in more recent times: the fact that 
$\rlam$ as estimated in the context of quantum field theories such as 
quantum chromodynamics (QCD), electroweak (EW) and grand unified theories 
(GUTs) implies impossibly large values of $\Olamo$ (Table~1).
These theories have been successful in the microscopic realm.
Here, however, they are in gross disagreement with the observed facts of
the macroscopic world, which tell us that $\Olamo$ cannot be much larger 
than order unity.  This ``cosmological constant problem,'' is undoubtedly
another reason why many cosmologists have preferred to set $\Lambda = 0$,
rather than deal with a parameter whose microphysical origins are still
unclear (see \cite{Car01} for review).

\renewcommand\arraystretch{1.15}
\begin{table}[t!]
\begin{center}
\footnotesize{Table 1.  Theoretical Estimates of $\Olamo$}
\begin{tabular}{|c|c|c|c|} \hline
Theory & \multicolumn{2}{c|}{Estimated value of $\rlam$} & $\Olamo$ \\ \hline
QCD & (0.3 GeV)$^4 \hbar^{-3} c^{-5}$ & $10^{16}$~g~cm$^{-3}$ & $10^{44}\hub^{-2}$ \\
EW & (200 GeV)$^4 \hbar^{-3} c^{-5}$ & $10^{26}$~g~cm$^{-3}$ & $10^{55}\hub^{-2}$ \\
GUTs & ($10^{19}$ GeV)$^4 \hbar^{-3} c^{-5}$ & $10^{93}$~g~cm$^{-3}$ & $10^{122}\hub^{-2}$ \\ \hline
\end{tabular}
\vspace{-3mm}
\end{center}
\end{table}
\renewcommand\arraystretch{1}

This, however, is no longer an appropriate response because observations
now indicate that $\Olamo$, while nowhere near the size suggested by
Table~1, is nevertheless greater than zero.  The cosmological constant
problem has therefore become {\em more\/} baffling, in that any quantum
field-theoretic account of this parameter must apparently contain
a cancellation mechanism which is not only good to some 44 (or 122)
decimal places, but which begins to fail at precisely the 45th (or 123rd).
One possibility is to treat $\Lambda$ as a dynamical quantity rather 
than a constant of nature, in which case the observed smallness of 
$\Olamo$ might be attributed to the {\em age\/} of the Universe 
\cite{Ove98}.  It is equivalent to introduce a fifth element
(known as {\em quintessence\/}) into cosmology \cite{Kra00}.
In general, however, this means extending Einstein's equations~(\ref{EFEs})
to incorporate new (and so far unobserved) phenomena such as scalar fields,
which in turn introduce new terms (intermediate to the $\Omato$- and 
$\Olamo$-terms) into Eq.~(\ref{modFE}).  While these ideas hold
theoretical promise, it is likely that they too must involve
fine-tuning if they are to reproduce exactly the values of $\Olamo$
observed.  As an alternative explanation, it has been proposed that
a universe in which $\Olamo$ was too large (or small) might be
incapable of giving rise to intelligent observers, so that the fact
of our own existence already ``requires'' $\Olamo\sim1$ \cite{Wei01}.

Let us pass now to the observational arguments for $\Olamo$.
Some have been mentioned already.  It was noted in \S\ref{sec:cos} (Fig.~4)
that numerical simulations of {\em large-scale structure formation\/}
match models with $\Olamo=0.7$ better than those with $\Olamo=0$.
Additional simulations would allow us to explore the parameter 
space with higher resolution.

Some of the arguments for {\em exotic CDM\/} in \S\ref{sec:exo} also show
a (more modest) dependence on $\Olamo$.  The trend in most cases is toward
higher values of $\Olamo$ in conjunction with lower values of $\Omato$.
In the arguments from cluster $M/L$-ratios and mass power spectra,
for example, we have seen that raising $\Olamo$ from zero to one
corresponds to a drop of $\sim40$\% in the preferred value of $\Omato$.

Tentative lower limits on $\Olamo$ have come from {\em galaxy number counts\/}.
The comoving volume is enhanced at large redshifts ($z\gtrsim 2$)
for high-$\Lambda$ models, leading to greater (projected) 
number densities at faint magnitudes.  In practice, it has proven
difficult to disentangle this effect from galaxy evolution.  Early claims 
of a best fit at $\Olamo\approx 0.9$ \cite{Fuk90} have been disputed on
the basis that the steep increase seen in numbers of blue galaxies
is not matched in the $K$-band \cite{Gar93}.  Attempts to account for
evolution in a comprehensive way have recently produced the 2$\sigma$
lower limit $\Olamo > 0.53$ \cite{Tot97}, with a good fit (for flat
models) at $\Olamo\approx 0.8$ \cite{Tot00}.  A preliminary plot of
constraints across the $\Omato-\Olamo$ plane \cite{Roc00} is consistent
with these results.

Measurements of $\Olamo$ from {\em gravitational lens statistics\/} 
are based on a similar premise: the increase in path length to a given
redshift in high-$\Lambda$ models should mean that more lensed sources are
seen.  Somewhat surprisingly, comparison of the observed frequency of
lensed quasars to that expected produces results in poor agreement with 
those inferred from galaxy counts; leading in fact to the strongest
current {\em upper\/} limit on vacuum density: $\Olamo < 0.66$
(2$\sigma$) for flat models \cite{Koc96b}.  Dust could obscure the distant
lenses and allow for a larger value of $\Olamo$ \cite{Mal97}.  This
objection can however be met by moving to radio lenses, which give an 
only slightly weaker 2$\sigma$ bound: $\Olamo < 0.73$ (for flat models)
or $\Olamo \lesssim 0.4 + 1.5\Omato$ (for nonflat ones) \cite{Fal98}.
Other sources of potential systematic error remain, including source
redshift distributions, survey incompleteness, lens modelling and
evolution in the mass profiles of the lensing objects.  A recent
2$\sigma$ limit from radio lenses is $\Olamo < 0.95$ \cite{Coo99}.

\begin{figure}[t!]
\begin{center}
\includegraphics[width=61.92mm]{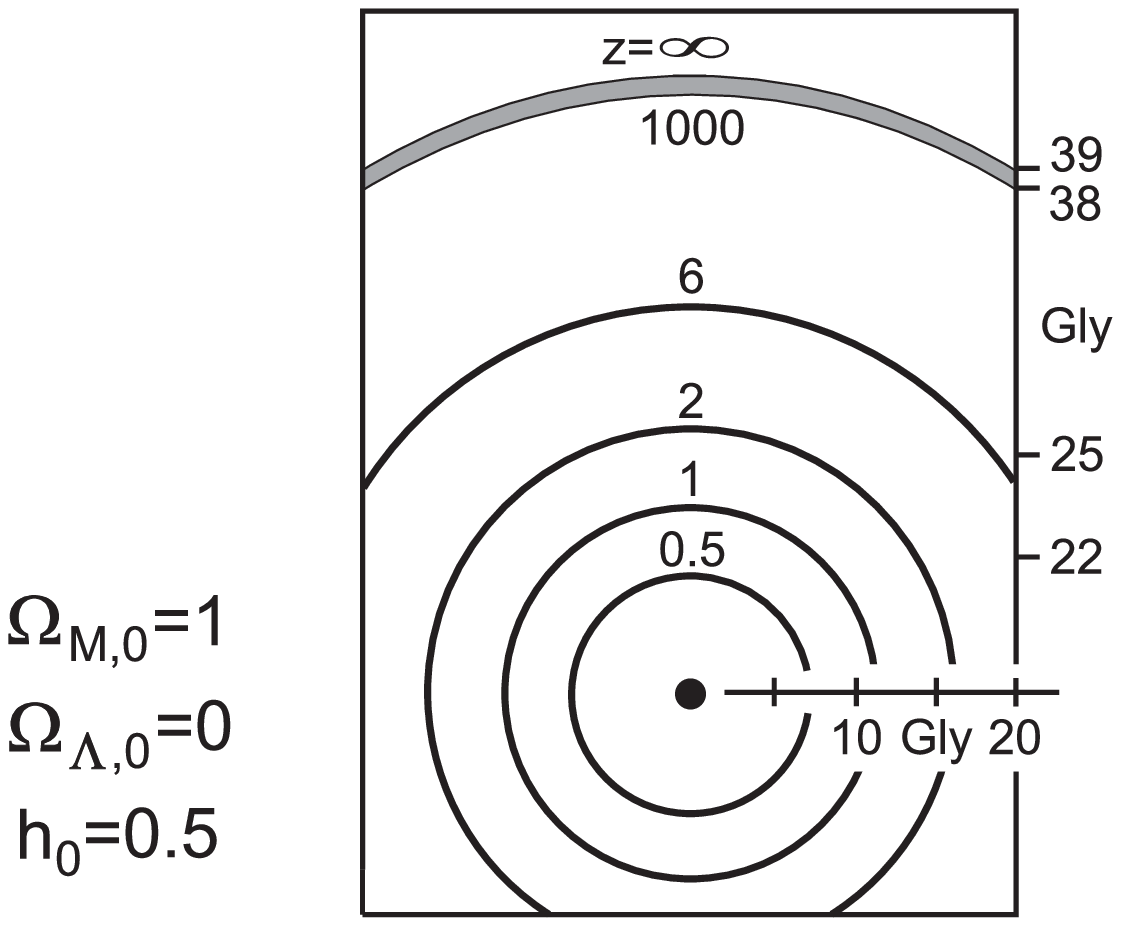}
\end{center}
\footnotesize{Fig.~6.  Two-dimensional slices in time of the flat EdS
              model ($\Omato=1,\Olamo=0$) with $\hub=0.5$ (left), 
              and a closed, low-$\Omato$, high-$\Olamo$ model 
              \cite{Lie92a,Lie92b} with $\hub=0.9$ (right).  
              There is nearly ten times as much linear distance
              between redshifts $z=3$ and 4 in the closed model as there
              is in the flat one.  (Both figures to same scale; 
              1~Gly$ = 10^9$~light-years.)}
\end{figure}

Lensing provides a second constraint on $\Olamo$ in closed models, based
on the {\em redshift of the antipodes\/} (the set of points at radial 
coordinate distance $\pi$).  It has been shown \cite{Got89} that the 
lensing cross section blows up for sources at this distance, implying 
that the antipodes cannot be nearer than the farthest normally lensed 
object, currently a galaxy at $z=4.92$ \cite{Fra97}.  It is straightforward
to compute the antipodal distance in terms of $\Omato$ and $\Olamo$; 
one finds in this way that $\Olamo<1.57$ if $\Omato=0.3$, an upper limit
that drops to $<1.30$ if $\Omato=0.1$.\footnote{A recent claim \cite{Epp00} 
   that this technique ``rules out values of $\Olamo$ between 1.0 and 1.5
   for $0 \leqslant \Omato \leqslant 0.3$'' is therefore incorrect.
   The appropriate upper limit (for matter densities in this range) is 
   $\Olamo < 1.5$, as reported in \cite{Whi96}.}
The upper limit is $\Olamo < 1.10$ if $\Omato=0.014$, essentially the same
as the Einstein limit discussed in \S\ref{sec:cos}.  Fig.~6 shows a 
two-dimensional slice of this model, along with the EdS model drawn to 
the same scale.  The antipodes (in the closed case) are at $z\approx 12$, 
well beyond the highest-redshift objects seen to date.  Fig.~6 also reveals
nearly ten times as much {\em linear\/} distance between redshifts $z=3$ 
and $z=4$ in the spherical model as in the flat one; this is why one 
expects to see more lensed sources (and faint galaxies) in a high-$\Lambda$
universe.

\begin{figure}[t!]
\begin{center}
\includegraphics[width=85.00mm]{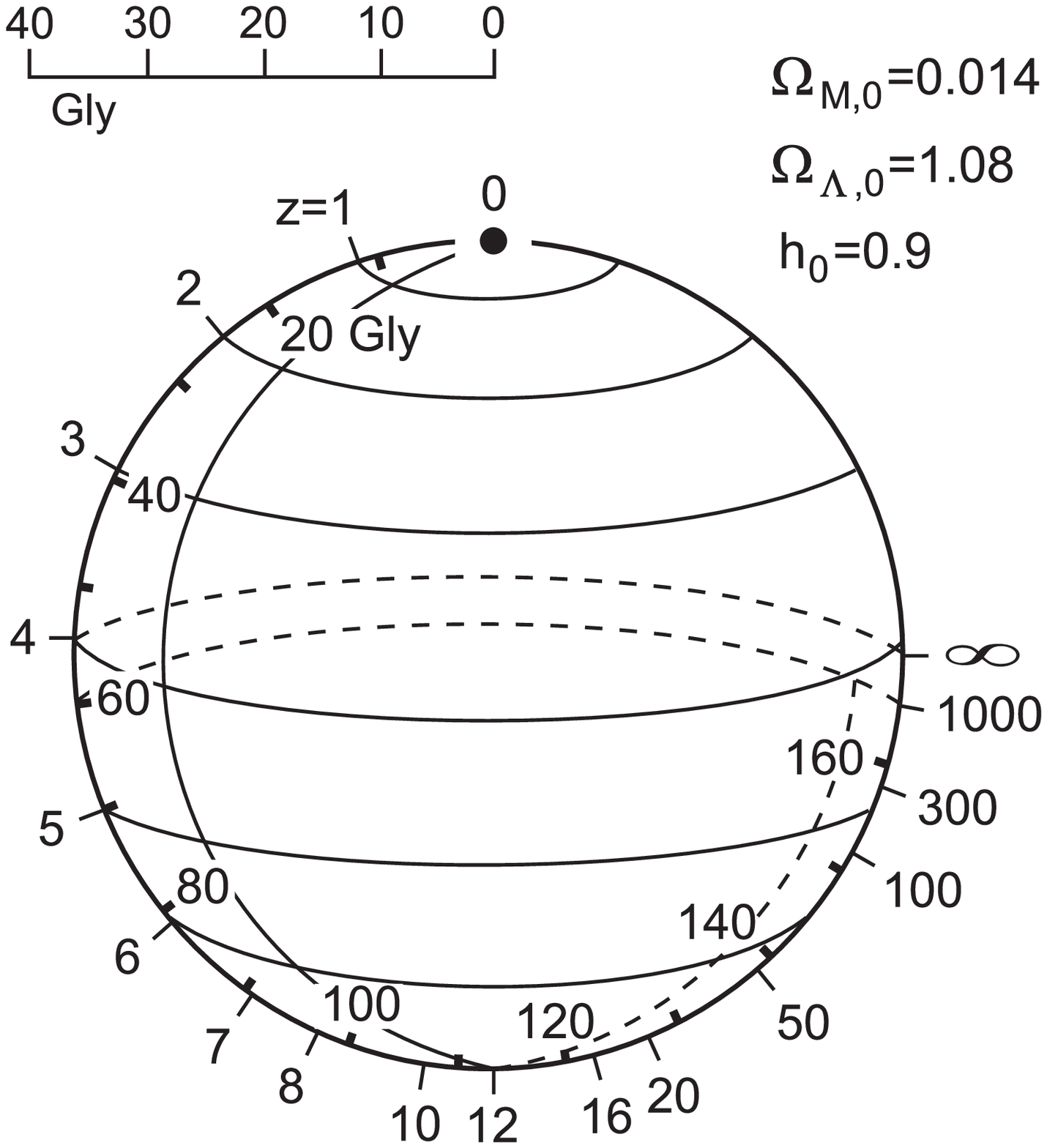}
\end{center}
\end{figure}

The best constraints on $\Olamo$ come from 
{\em Type Ia supernovae\/} (SNe~Ia), used as standard candles
in the classical magnitude-redshift relation, can measure $\Olamo$ more 
robustly than either galaxy counts or lensing statistics because evolution
is less of a concern for these objects.  The Supernova Cosmology Project
\cite{Per99} and High-$z$ Supernova Search teams \cite{Rie98}
have lately caused a great stir with their reports that a systematic
dimming of SNe~Ia at $z\sim0.5$ by about 0.25~mag relative
to that expected in an EdS model is best explained by
\beq
\Olamo \approx \frac{4}{3} \, \Omato + \frac{1}{3} \pm \frac{1}{2} \; \; \;
   (2\sigma) \; .
\label{SNeIa}
\eeq
This leads at once to a conclusion that most cosmologists\footnote{With some
   exceptions \cite{Efs90,Lie92a,Lie92b}; see \S\ref{sec:bnp}.}
have long been reluctant to consider: that 
{\em we live in a vacuum-dominated universe\/}.
Several words of caution are in order, however.
Observations must reach $z\sim2$ before one can truly be certain of 
discriminating between models like those shown in Fig.~2.
Intergalactic dust could mimic the effects of a cosmological constant,
if it were ``sifted'' during the process of ejection from galaxies
\cite{Agu00}.  The neglect of evolution may be more serious than 
claimed \cite{Dre00}.  And the physics of SNe~Ia explosions needs 
to be better understood \cite{Hil00}.

The best constraints on $\Olamo$ come from 
{\em CMB fluctuations\/}; and in particular, from the new detection
of the first peak in their angular power spectrum by the 
{\sc Boomerang} \cite{deB00} and {\sc Maxima} \cite{Han00} experiments.
The location of this peak is a direct measure of the largest
size of fluctuations in the primordial plasma at the moment of last 
scattering, as seen through the ``lens'' of a curved Universe.
The two detections, combined with earlier data from the {\sc Cobe} 
satellite \cite{Ben96}, imply that \cite{Jaf00}
\beq
\Olamo = 1.11^{+0.13}_{-0.12} - \Omato \; \; \; (2\sigma) \; .
\label{CMB}
\eeq
This result is more reliable than all the others discussed so far
because it bypasses ``local'' phenomena such as supernovae,
galaxies, and even lensed quasars; taking us directly back to the
radiation-dominated era when physics was very simple.  

Let us therefore use Eq.~(\ref{CMB}) to calculate the energy density
of the vacuum.  Summing the baryon, exotic CDM, and neutrino densities 
--- Eqs.~(\ref{Obar}), (\ref{OCDM}) and (\ref{Onu}) respectively --- 
gives the total matter density $\Omato$.  Substituting this 
into Eq.~(\ref{CMB}), we find the following ranges of values:
\beq
\Olamo = \left\{ \begin{array}{cc}
   0.3 - 1.1 & \mbox{ (GI theory) } \\
   0.4 - 1.2   & \mbox{ (otherwise) } 
\end{array} \right.
\label{Olam}
\eeq
Vacuum energy, the new and invisible ``fire'' of modern cosmology,
is thus indeed dominant, making up the bulk of a dark Universe in which
light and baryons --- the constituents of our familiar world ---
appear almost incidental.

\section{How much exotic dark matter?} \label{sec:bnp}

The basis for $\Lambda$CDM as the new favorite among cosmological
models lies in the approximate orthogonality of the CMB and supernova 
bounds, Eqs.~(\ref{SNeIa}) and (\ref{CMB}).  Indeed, if we take 
{\em both\/} of these results at face value, we can substitute
one into the other and solve to find $\Olamo = 0.78 \pm 0.23$ and
$\Omato = 0.33 \pm 0.22$.  The fact that this latter number is very
near the center of the range of allowed values for $\Omato$ in 
Eq.~(\ref{OCDM}) has then been taken as a further sign of the
basic correctness of both the $\Lambda$CDM model in particular and
the GI theory of structure formation in general.

While this is a self-consistent account, and one that agrees with most
observations, it suffers from one flaw: it is inherently improbable.  The
densities of baryonic matter (and exotic CDM, should it exist) evolve at a
very different rate from neutrinos; and both of these components evolve at
very different rates from vacuum energy.  So one has three kinds of matter
which should not have anything like the same density parameters at any given
time --- and yet two of them (at least) do.  In the $\Lambda$CDM picture, in 
particular, it seems that we happen to live at a time when $\Olamo$ and
$\Omato$ are separated by a mere factor of two.  To illustrate the
unlikelihood of this ``preposterous universe,'' Carroll \cite{Car01}
has plotted the evolution of $\Omat$ and $\Olam$ for 35 powers of ten
in scale factor in the past and future directions, showing that the
probability of finding oneself at the moment when they should be
within even one order of magnitude of each other is exceedingly
remote.

It may be misleading to characterize this problem in logarithmic terms.
Several Gyr were necessary to form the first galaxies and stars; and they
will all be gone after a hundred --- so it is natural that we should
find ourselves within this span of cosmic history, at least.
In Fig.~7, we have plotted the evolution of $\Olam(t)$ and $\Omat(t)$ 
on a {\em linear\/} scale for the first $\sim80$ (100)~Gyr after the 
big bang in the $\Lambda$CDM model with $\hub=0.9$ (0.7).
This plot confirms that $\Lambda$CDM is improbable, in the sense that
the values of $\Omat$ and $\Olam$ observed at the present epoch
(0.3 and 0.7 respectively) are atypical. 

Fig.~7 also shows the evolution of $\Olam$ and $\Omat$ in the 
vacuum-dominated, $\Lambda+$baryon model discussed in several places
above (Model~1 in Figs.~2 and 3), with $\Olamo=1.08$ and no exotic CDM
($\Omato=0.014\approx\Obar$).  This universe, which was originally proposed
in \cite{Lie92a,Lie92b} and which we shall term here the ``\Lbar''model,
is a good deal less preposterous than $\Lambda$CDM in the sense that 
a factor of $\sim80$ (rather than two) separates the presently observed
values of vacuum and energy density.  Indeed these parameters are much closer
to their ``cosmological average'' values of one and zero respectively.
While this in itself does not constitute a case for the model,
it prompts us to wonder whether $\Lambda$ might not be more important
than most cosmologists have been willing to consider.
Could vacuum energy be not just the dominant, but the 
{\em only significant component of the dark matter?\/}

\begin{figure}[t!]
\begin{center}
\includegraphics[width=75mm]{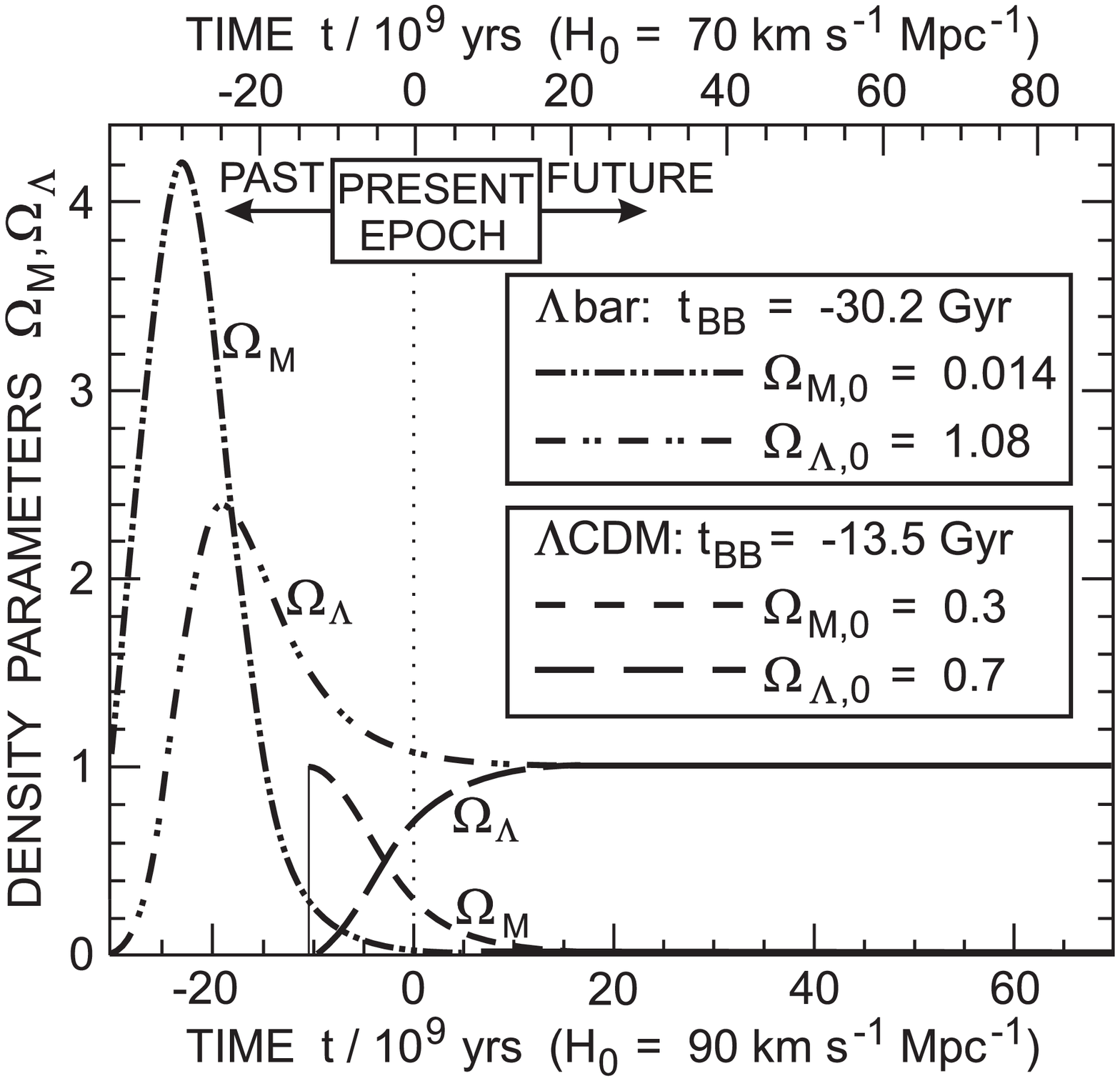}
\end{center}
\footnotesize{Fig.~7.  Evolution of $\Omat$ and $\Olam$ in the $\Lambda$bar
              and $\Lambda$CDM models (Models~1 and 6 in Figs.~2 and 3).
              Time is set to zero at the present epoch; $\bs{t}{BB}$, the
              time of the big bang, is calculated using $\hub=0.9$ for
              $\Lambda$bar (bottom scale) and $\hub=0.7$ for $\Lambda$CDM
              (top scale).  Compare Fig.~3.}
\label{prep}
\end{figure}

Such an idea would have been unthinkable only a few years ago, when
it was still routine to set $\Lambda=0$ and cosmologists had two main 
choices: the ``one true faith'' ($\Omato\equiv1$), or the ``reformed''
(with each believer being free to choose his or her own value
near $\Omato\approx0.3$).  All this has been irrevocably altered by 
the CMB experiments (\S\ref{sec:vac}).
If there is a single guiding principle in choosing models now,
it is no longer $\Omato\approx0.3$, and certainly not $\Olamo=0$;
it is $\Omato+\Olamo\approx1$ from the power spectrum of the CMB.
With this in mind, we will devote this final section of our review
to exploring the feasibility of the \Lbar\ model, which has 
$\Omato+\Olamo=1.094$, in excellent agreement with Eq.~(\ref{CMB}).

To begin with, any model of this kind, with a small $\Omato$ and a high
$\Olamo$, must face three main objections:
(1)~the lower limits on $\Omato$ in \S\ref{sec:exo}; (2)~the upper limits
on $\Olamo$ in \S\ref{sec:vac}; and (3)~the age problem in \S\ref{sec:bar}.
Let us briefly review these points before moving on to what are, to us,
the strongest arguments in {\em favor\/} of the \Lbar\ idea.

We have argued in \S\ref{sec:exo} that the lower limits on $\Omato$ are 
of two kinds: those which are framed in the context of gravitational 
instability theory (thus tacitly assuming the existence of large amounts
of CDM) and those which are not.  Of these, the former are certainly 
incompatible with $\Omato=0.014$.  The latter, however, are more 
flexible, especially when their $\Olamo$- and $\hub$-dependence
is taken into account (as it often is not).  The distinction is important,
because CDM may not be {\em needed\/} in the \Lbar\ model:
structures have far longer to form, and they do so at a time when
expansion is slower and densities are higher.

The observational constraints on $\Olamo$ in \S\ref{sec:vac} are
not much more secure than those on $\Omato$.  Supernovae favor a value 
of $\Olamo$ closer to $\sim0.8$, but have large (and possibly underestimated)
uncertainty factors.  The antipodal redshift argument restricts us to 
$\Olamo<1.10$, which is just above the value we have adopted --- 
suggesting that this could become a useful test of the theory when 
observations are eventually pushed to $z\approx 12$ (Fig.~6).
Gravitational lensing statistics remain the strongest argument against
values of $\Olamo$ as large as that considered here.  It would be a more
compelling one, however, if we understood why galaxy counts
(which rely on essentially the same reasoning) lead to such
different conclusions.

Finally, as we have argued in \S\ref{sec:cos} and \S\ref{sec:bar}, the
age of the \Lbar\ universe should be seen as an asset rather than a 
liability, particularly in the area of structure formation.  One also 
expects to find a faint population of very old, dead stars; these may be
the halo objects whose existence is implied by the {\sc Macho} survey.
The most direct way to rule out the \Lbar\ model based solely on age 
considerations would be to prove that $\hub$ is less than or equal 
to what we have referred to as the ``low'' value (0.7) above.
Alternatively, if evidence continues to mount for the ``high'' value
(0.9), then the ``age problem'' will begin to put pressure on the
$\Lambda$CDM, rather than the the \Lbar\ model.

Let us turn now to the observational case for the \Lbar\ idea,
which rests on two main lines of evidence: Ly$\alpha$ absorption spectra,
and the lack of a {\em second\/} peak in the power spectrum of 
CMB fluctuations.  We consider these in turn.

The forest of Ly$\alpha$ absorption lines was first used to measure 
$\Omato$ and $\Olamo$ in \cite{Lie92a,Lie92b}; the most recent review 
of this method appears in \cite{van99}.
The idea is extremely simple: one supposes that Ly$\alpha$ absorbers,
like galaxies, are distributed with a cell-like structure, and that
absorption lines are produced when the line of sight to a distant
quasar cuts through the cell walls (Fig.~8).
The crucial assumption is then made that the cells expand with
the Hubble flow, and that evolution {\em within\/} them is secondary and
can be neglected.  This is not unreasonable, given that the expansion 
velocity of a typical cell would be of order $\sim3000$~km/s, whereas
peculiar motions inside the cell walls might be no more than about
$\sim300$~km/s.  This picture is also in accord with the latest observational
work on Ly$\alpha$ absorbers which, although still tentative, suggests that
they are indeed distributed in structures which expand with
decreasing redshift \cite{Din98}, and have comoving size 
$\sim26 \hub^{-1}$~Mpc at $z=2.6$ \cite{Wil00b}.\footnote{In
   fact, neither assumption may be strictly necessary; a similar analysis
   has been performed under the assumption that Ly$\alpha$ absorbers make up
   a homogeneous population of clouds with constant comoving size, and it
   leads to results consistent with those presented here \cite{Hoe94}.}

\begin{figure}[t!]
\begin{center}
\includegraphics[width=85mm]{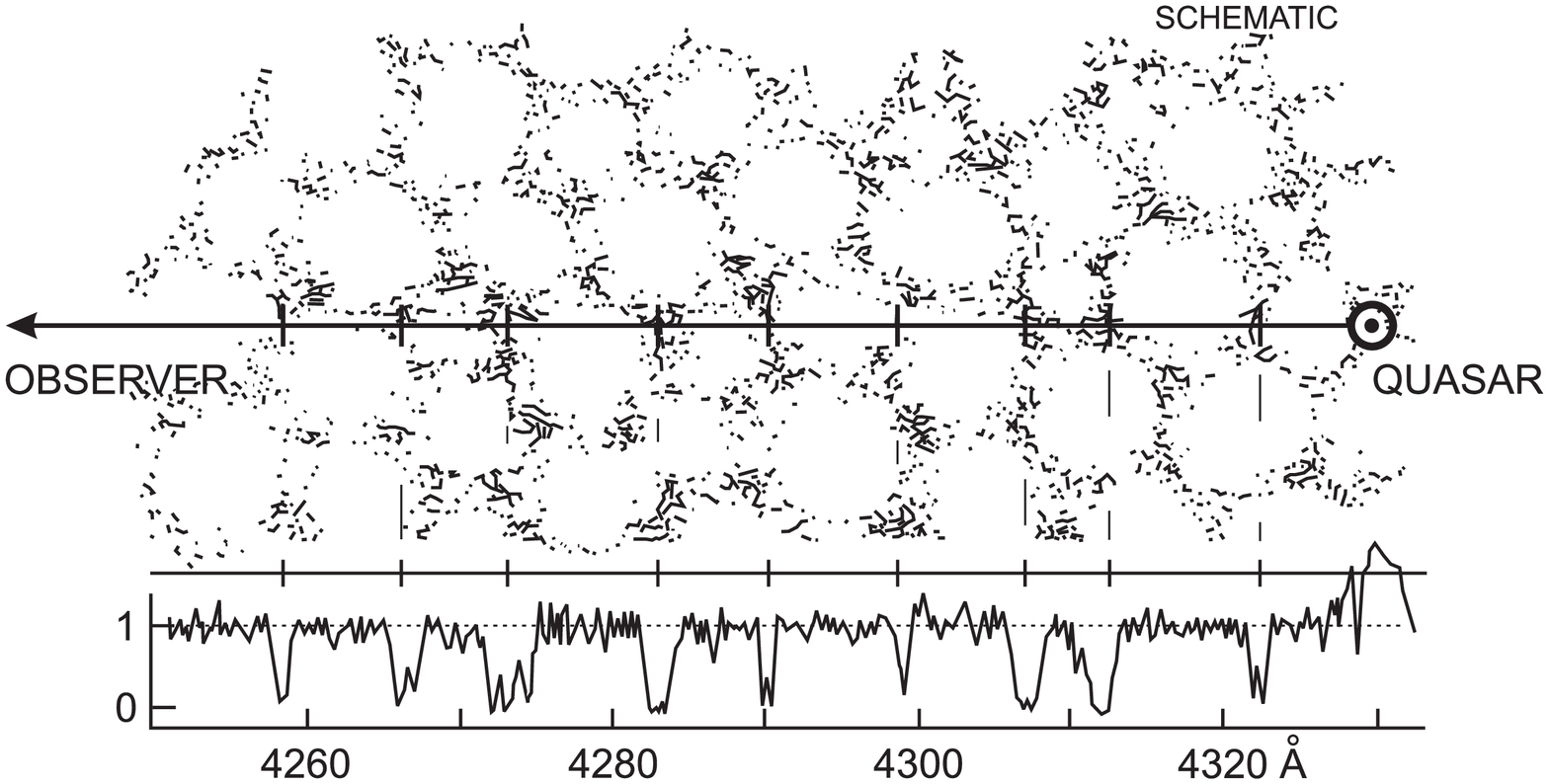}
\end{center}
\footnotesize{Fig.~8.  Schematic illustration of a network of cell-like
              structures, with Ly$\alpha$ absorption lines seen by an
              observer in the spectrum of a distant quasar due to the 
              fact that the line of sight passes through the cell walls.}
\end{figure}

One then simply counts the absorption lines and measures the mean spacing 
$\Delta\lambda$ between them.  This gives the redshift spacing
$\Delta z(z)$ of the cells as a function of redshift, which may in turn 
be related to their comoving coordinate size $\Delta\chi$ by
\beq
[\Delta z(z)]^2 = \left( \frac{\bs{R}{0} \Delta\chi}{c} \right)^2 
   \!\!\! H^2(z) \; ,
\label{poly}
\eeq
where $\chi$ is the radial coordinate distance and $H^2(z)$ is given as a
cubic in $(1+z)$ by Eq.~(\ref{FE}).  Since $\Delta z(z)$ is an observable, 
Eq.~(\ref{poly}) becomes a third-order polynomial regression formula 
for $[\Delta z(z)]^2$; and one, moreover, with {\em no linear term\/}.

Fig.~9 shows a plot of $[\Delta z(z)]^2$ versus $(1+z)$, based on published
spectra of 21 different quasars with a total of 1320 Ly$\alpha$ absorption
lines, weighted according to resolution (see \cite{Lie92b} for details).
At first sight it does not seem that the regression curve is strongly
constrained by the data.  However, the fit is in fact remarkably robust.
The reason for this is that the curve can consist of only three
components: a constant, a downward-opening quadratic, and a cubic 
originating at (0,0).  The regression coefficients that meet these
conditions span only a very narrow range of values, and lead directly
to the 2$\sigma$ (statistical) results \cite{Lie92b}
\beqa
\Omato & = & 0.014 \pm 0.006 \nonumber \\
\Olamo & = & 1.08 \pm 0.03 \; .
\label{BNP}
\eeqa
Eqs.~(\ref{BNP}) define what we have referred to as
the \Lbar\ model; this term could equally be applied to other
vacuum-dominated models without significant amounts of CDM 
(e.g., Fig.~10).

The \Lbar\ model presented in Eqs.~(\ref{BNP}) passes several basic
consistency tests.  Firstly, the sum of $\Omato+\Olamo$ matches that 
seen in the CMB experiments (\S\ref{sec:vac}).
Secondly, the value of $\Omato$ is within the bounds imposed by cosmic
nucleosynthesis (\S\ref{sec:bar}), favoring low values of $\Obar$ and 
high values of $\hub$ ($\Obar\leqslant0.016\hub^{-2}$ for $\hub=0.9$,
or $\hub\geqslant0.71$ for $\Obar=0.01\hub^{-2}$).
And thirdly, the regression curve (heavy solid line in Fig.~9) passes 
through $z=0$ at $\Delta z\approx 0.009$, in excellent accord with
the distribution of galaxy structure seen in our own cosmic neighborhood
\cite{Gel89} (the empty rectangle in the upper left-hand
corner of Fig.~9).  These phenomena involve independent physics on 
widely different scales, and we would regard it as remarkable for a 
simple procedure like the one described above to agree with all three
by chance alone.

\begin{figure}[t!]
\begin{center}
\includegraphics[width=85mm]{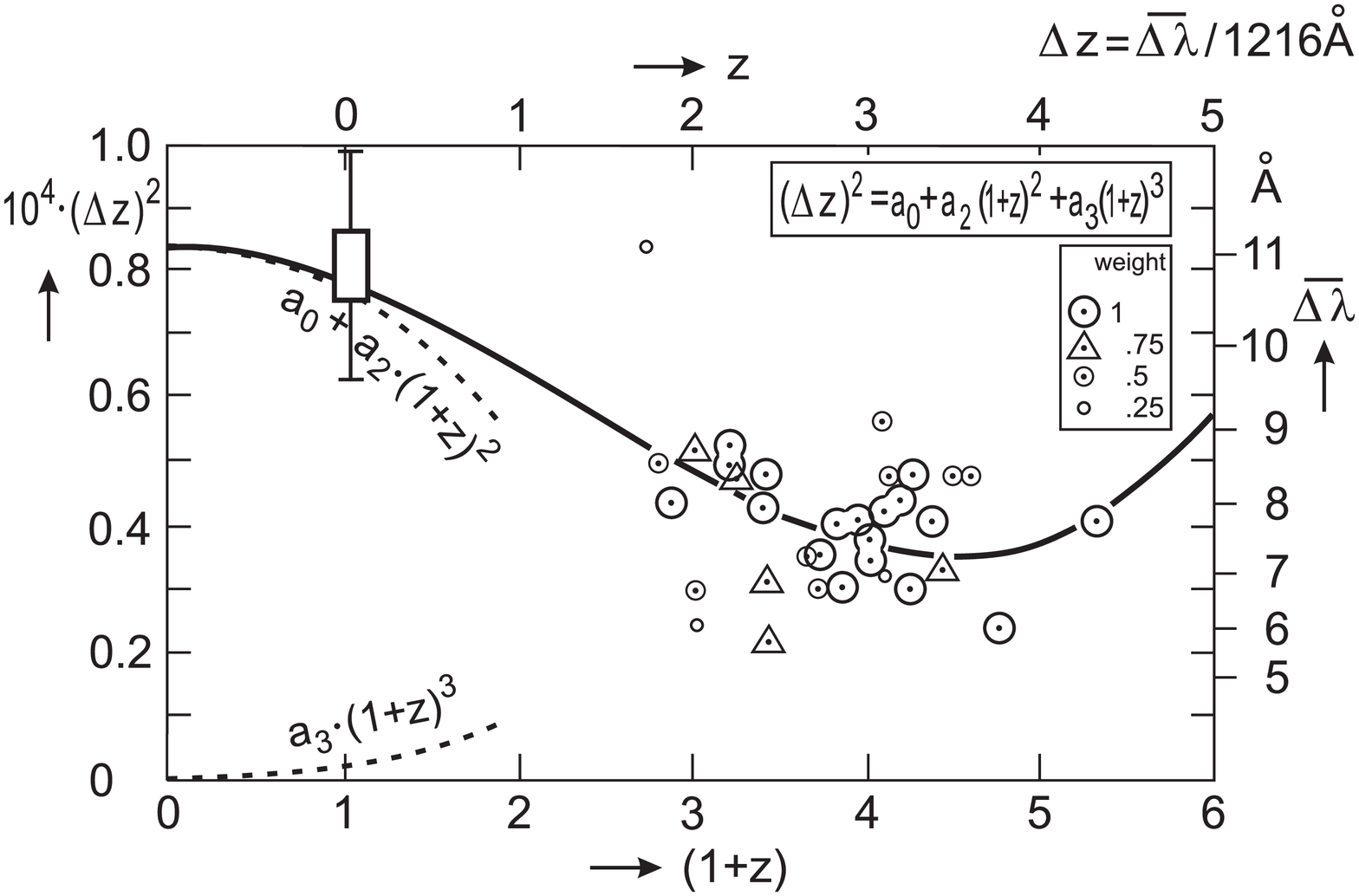}
\end{center}
\footnotesize{Fig.~9.  Best fit curve of $\Delta z(z)$ to the data from
              Ly$\alpha$ absorption lines in the redshift range
              $1.8 \leqslant z \leqslant 4.5$.  Data are weighted according
              to the quality of the spectra.  Dashed lines show the quadratic
              and cubic components of the fit.  The coefficients $\bs{a}{0},
              \bs{a}{2}$ and $\bs{a}{3}$ are proportional to $\Olamo$,
              $(1-\Ototo)$ and $\Omato$ respectively.}
\end{figure}

The third point however deserves some clarification.  The fact that the
scale of the local galaxy distribution matches that of Ly$\alpha$ absorbers
agrees with our simple picture, in which both types of matter cluster 
predominantly along the cell walls (Fig.~8). 
However, there is still significant debate in the literature over the 
extent of correlations between Ly$\alpha$ absorbers and other 
structures \cite{Rau98}.  Why not ``calibrate'' the regression 
curve at $z=0$ using absorbers rather than visible galaxies?
Ultraviolet Ly$\alpha$ spectra are now available from the Hubble Space
Telescope ({\sc Hst}) over the redshift range $0 \leqslant z \leqslant 1.3$
\cite{Bah96}.  The absorber material, however, is not spread uniformly over
the cell walls, but rather occurs in filaments and knots.  Lines of sight
to a few quasars, therefore, are far less likely to ``notice'' the cell walls
at low redshifts, where cell size is enormous.  Indeed, the {\sc Hst}
spectra show a mean redshift spacing $<\!\!\Delta z\!\!> \approx 0.04$,
four times larger than that seen in the galaxy distribution \cite{Gel89}.
In the local universe, in other words, galaxy surveys are better guides
to large-scale structure than Ly$\alpha$ absorption spectra.

A different restriction comes into play at {\em high\/} redshifts,
where one might expect from the above argument that Ly$\alpha$ spectra
would be increasingly reliable.  It is certainly true that lines of sight
to quasars at $z \gtrsim 4$ will ``notice'' most of the cell walls they
intersect, since the cells themselves are smaller (by a factor $1+z$) 
at these redshifts.  Another effect, however, also goes as $1+z$:
Doppler broadening of lines due to peculiar motions of the Ly$\alpha$ 
absorbers in the cell walls.  At $z\approx 4$ the latter will already 
be of order $\sim$1500~km/s rather than $\sim$300~km/s, introducing
spurious lines which would masquerade as small-scale structure.
Recent Keck/{\sc Hires} spectra of a $z=4.1$ quasar confirm this 
suspicion, yielding a mean redshift spacing
$<\!\!\Delta z\!\!> \approx 0.002$ \cite{Lu96}, three times 
{\em smaller\/} than the fit to the regression curve at $z\approx4$
(Fig.~9).  The Ly$\alpha$ method outlined here is most sensitive to 
the cosmological parameters for quasars in the redshift range
$2 < z < 4$.  The most recent Keck/{\sc Hires} spectra of a quasar at 
$z=3.1$, for example, shows $<\!\!\Delta z\!\!> \approx 0.006$ \cite{Kir97},
in excellent agreement with Fig.~9 at $z\approx 3$.

We have not addressed the possibility of bias due to inherent
evolution within the cell structure.  There is still debate about the mean
comoving scale of the Ly$\alpha$ distribution in the literature, to say 
nothing of its possible time rate of change.  To obtain higher values of 
$\Omato$ from the above analysis, however, evolution must be in the right
direction.  An appeal to ionization, for instance, should not ionize 
high-redshift clouds more than local ones; this would raise $\Delta z$ 
at high $z$ and {\em lower\/} the inferred value of $\Omato$.  Higher 
matter densities can in principle be obtained if evolution is such that 
substructure increases with redshift.  Observations and numerical 
simulations do suggest the possibility of such a trend in the 
Ly$\alpha$ forest, as sheet-like structures give way to filaments 
and knots with time \cite{Rau98}.  It has been estimated that 
allowance for an effect of this kind might raise the value of 
$\Omato$ from that in Eq.~(\ref{BNP}) to as much as $\sim0.05$
\cite{van99}.

Questions of a more mathematical nature may be raised by the smallness
of the uncertainties in Eqs.~(\ref{BNP}).  The possible impact of spectral
resolution, equivalent width, and line blending on the line-counting
procedure have been considered in \cite{Lie92b}, with the conclusion that
corrections arising from these factors will be minor.  The statistical 
robustness of the Ly$\alpha$ method can be attributed to the absence 
of a linear term in Eq.~(\ref{FE}) for Hubble's parameter.
More work, however, could be done to reduce the possibility
of unmodelled systematic errors.  Ultimately one would like to
see $\Delta z(z)$ extracted from a power spectrum analysis using high
signal-to-noise spectra (like those now coming from Keck/{\sc Hires})
together with automated line-fitting and counting procedures \cite{Kir97}.

We move now to the second argument for a \Lbar\-type universe, one which 
relies on a new analysis of the angular power spectrum of the CMB (Fig.~10).
While the angular {\em positions\/} of the peaks in this spectrum fix 
the sum of matter and vacuum densities (\S\ref{sec:vac}), their relative
{\em heights\/} are largely a function of the matter density alone.
The odd-numbered peaks are produced by regions of 
the primordial plasma which have been maximally compressed by infalling
material, while the even ones correspond to maximally rarefied regions 
which have rebounded due to photon pressure. 
A high baryon-to-photon ratio enhances the compressions and retards the 
rarefractions, thus suppressing the size of, e.g., the second peak 
relative to the first.  The strength of this effect depends on the 
fraction of baryons (relative to the more weakly-bound neutrinos and
CDM particles) in the overdense regions.

\begin{figure}[t!]
\begin{center}
\includegraphics[width=85mm]{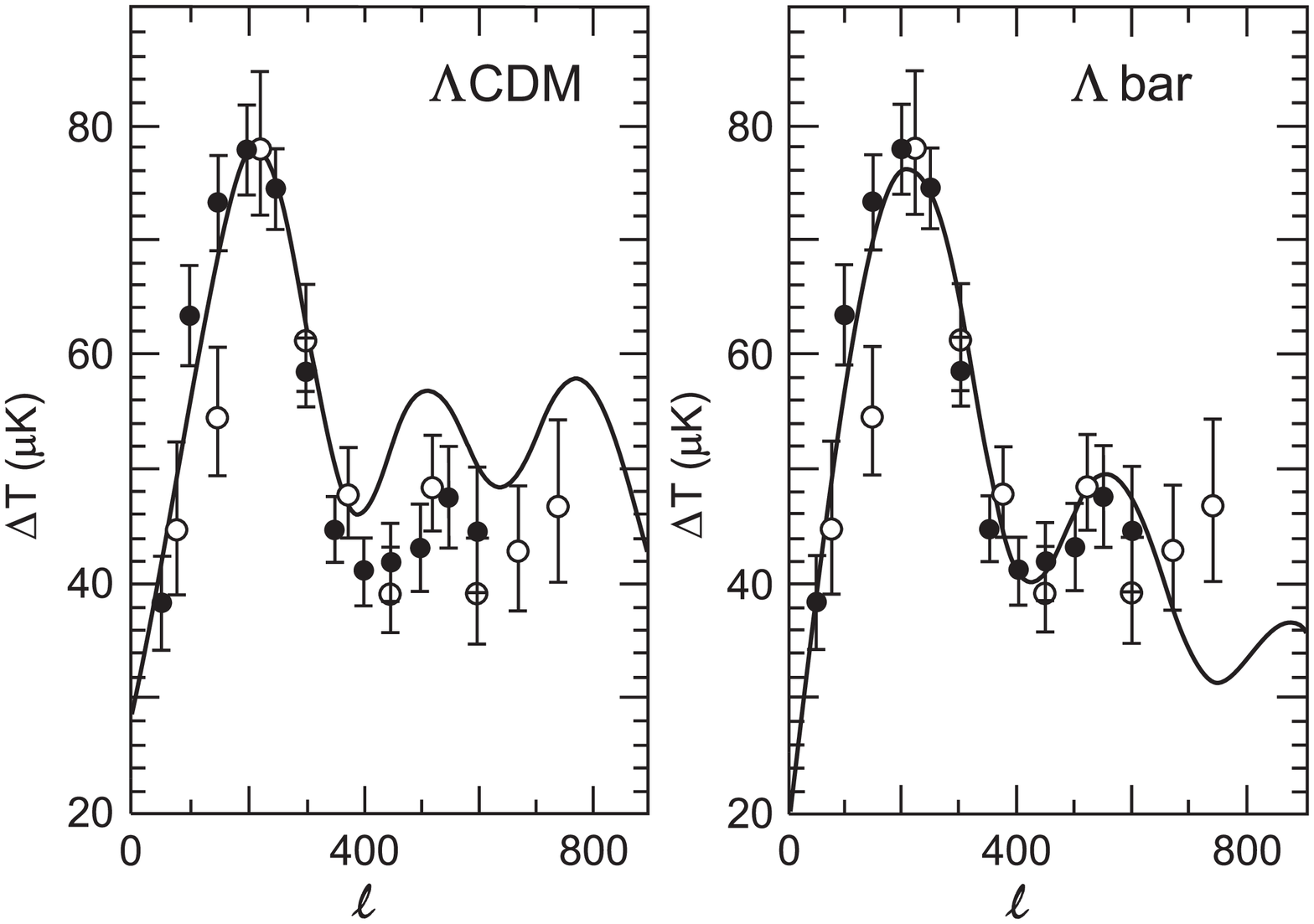}
\end{center}
\footnotesize{Fig.~10.  Observational data on the CMB power spectrum
              from the {\sc Boomerang} (filled circles) and 
              {\sc Maxima} experiments (open circles), together with
              theoretical expectations based on a $\Lambda$CDM model 
              (left) with $\Obar=0.039, \Ocdm=0.317, \Olamo=0.644$ and 
              $\hub=0.7$; and a closed \Lbar\ model (right) with 
              $\Obar=0.034, \Ocdm=0, \Olamo=1.006$ and $\hub=0.75$.
              Figure courtesy S.~McGaugh \cite{McG01}.}
\end{figure}

Data taken by the {\sc Boomerang} and {\sc Maxima} experiments appear to 
show an almost {\em total\/} suppression of the second peak relative to the 
first, inconsistent (at the 99\% level) with expectations based on the 
$\Lambda$CDM model (Fig.~10, left-hand side).  The ratio of baryons to CDM 
in the primordial plasma, therefore, appears to be higher than predicted.
A first reaction might be to keep the CDM and raise the baryon density;
this however brings the theory into immediate conflict with nucleosynthesis
limits on $\Obar$ \cite{Teg00}.  Other remedies (within the framework of 
gravitational instability theory) include tilting the spectrum of initial 
perturbations to disfavor smaller-scale (higher-order) peaks \cite{Jaf00},
erasing these peaks outright with processes such as delayed recombination
\cite{Pee00b} or decoherence, and varying one or more constants of
nature \cite{Whi00}.

The alternative is to take the apparent lack of CDM at face value.
This can either be done in a half-hearted or whole-hearted way.  The 
half-hearted way is to retain a minimum density of CDM with a statistical 
``prior.''  Thus, requiring that $\Ocdm > 0.1$, but otherwise fitting 
the combined {\sc Boomerang, Maxima} and {\sc Cobe} data, one obtains 
a model with best-fit parameters $\Obar=0.032 \hub^{-2}$ and 
$\Ocdm=0.14 \hub^{-2}$ \cite{Jaf00}.

The whole-hearted approach, which may however require extending the 
standard picture of structure formation, is to drop the requirement of
CDM altogether.  Results are shown in Fig.~10 (right-hand side), 
which is a statistical fit to $\Obar$ with $\Ocdm=0$ and $\hub=0.75$ 
\cite{McG01}.  The values of $\Olamo$ and $\Delta T$ (amplitude) are 
fixed by the position and height of the first peak.  The best-fit 
model passes neatly through both peaks ($\chi^2=0.85$) and has 
\beqa
\Omato & = & \Obar \; \; = \; \; 0.034 \nonumber \\
\Olamo & = & 1.006 \; , 
\eeqa
in good agreement with primordial nucleosynthesis (\S\ref{sec:bar})
as well as Eqs.~(\ref{BNP}).  This model has an age of 22.2~Gyr (with 
$\hub=0.75$), and a total density parameter slightly above one,
in agreement with suggestions from other new analyses of the 
{\sc Boomerang} and {\sc Maxima} data \cite{Gri01,Whi00}.
The shape of the CMB power spectrum, along with the analysis of
Ly$\alpha$ absorption lines presented above, thus favors an old,
closed \Lbar\ model, dominated by vacuum energy, 
with no significant contributions from CDM or neutrinos --- 
a universe composed almost exclusively of ``fire'' and ``earth.''

\section{Conclusions} \label{sec:con}

We have reviewed the evidence for the ``four elements'' of 
modern cosmology: baryons (``earth''), exotic cold dark matter (``water''),
neutrinos and photons (``air'') and vacuum energy (``fire'').  Recalling
Eqs.~(\ref{Obar}), (\ref{OCDM}), (\ref{Onu}), (\ref{CMB}) and (\ref{Olam}),
we may summarize the present contributions  of each element to the total
density of the Universe with the following ranges of values:
\beqa
\Obar & = & 0.012 - 0.041 \nonumber \\
\Ocdm & = & \left\{ \begin{array}{cc}
   0.1 - 0.5 & \mbox{ (GI theory) } \\
   0 - 0.4   & \mbox{ (otherwise) } 
   \end{array} \right. \nonumber \\
\Onu & = & 0.0003 - 0.2\Ocdm \nonumber \\
\Olamo & = & \left\{ \begin{array}{cc}
   0.3 - 1.1 & \mbox{ (GI theory) } \\
   0.4 - 1.2   & \mbox{ (otherwise) } 
   \end{array} \right. \nonumber \\
\overline{\hspace{1cm}} \nonumber \\
\Ototo & = & 0.99 - 1.24 \; \; \; \mbox{\cite{Jaf00}} \; ,
\eeqa
where ``GI'' refers to gravitational instability theory.
Baryons, the stuff of which we are made, 
are apparently little more than a cosmic afterthought.
Neutrinos and the elusive cold dark matter may both play 
far more significant roles in determining the past and future
evolution of the Universe, though this is not certain.
What is clear is that all three forms of matter are dwarfed
in importance by a newcomer whose physical origin remains 
shrouded in obscurity: the energy of the vacuum.

We have devoted the last part of our review to the hypothesis 
that vacuum energy dominates so completely that there is no room
for significant amounts of neutrino or exotic dark matter at all.
This would considerably simplify our picture of the Universe,
ease problems with the ``preposterously'' fine-tuned values of
the observed cosmological parameters, and allow more time for 
galaxies and other structures to form.  It would also, however, 
require that we modify the GI paradigm.  We have reviewed the 
various lines of observational argument, both for and against such
an idea.  It appears to us quite possible that the vacuum density
$\Olamo$ is close to one, that the sole contributions to the
matter density $\Omato$ come from a small amount of baryons and 
neutrinos, and that $\Olamo$ and $\Omato$ together are enough to
``just close'' the Universe.

\vspace{15mm}

\renewcommand{\baselinestretch}{0.85}\normalsize
{\footnotesize {\em Acknowledgments.\/}
The authors are grateful to D.~Clowe, K.~S.~de~Boer, J.~Ehlers, P.~Schneider
and C.~van~de~Bruck for comments, and thank W.~Fusshoeller for technical 
assistance.  J.O. is supported by an Alexander von Humboldt Postdoctoral
Fellowship.}

\vspace{5mm}

\section*{Appendix: Axions and WIMPs}

\renewcommand{\baselinestretch}{1.0}\normalsize
{\small 
This appendix is a brief introduction to axions and supersymmetric
weakly interacting particles (WIMPs); and in particular, to their main
claim as compelling cold dark matter (CDM) candidates:
the theoretical expectation that either one, if it exists,
naturally has a collective density $\Ocdm$ close to the critical density.  

Axions come about as part of the solution to the ``strong CP problem'' in 
quantum chromodynamics (QCD).  This theory has had some enormous successes,
notably in accounting for quark confinement and asymptotic freedom.
However, it contains a dimensionless free parameter $\Theta$ which 
must be less than $\sim10^{-10}$ in order to explain why strong 
interactions do not violate parity (P) or charge-parity (CP).
(Upper bounds on the electric dipole moment of the neutron tightly constrain
any such violations.)  To remove this unnaturally small number from the
theory, particle theorists suppose that $\Theta$ is driven toward zero by the 
spontaneous breaking of a new symmetry of nature at energy scales $f_a$.
The axion ($a$) is the new boson which appears as a result of this 
phase transition, eventually acquiring a rest mass $m_a c^2 \propto 1/f_a$.
Unfortunately neither parameter is constrained {\em a priori\/} by theory
(although one hopes that $f_a$ is less than $10^{10}$ times the QCD phase
transition energy, or little would be gained from the whole mechanism).
Observation, however, comes to the rescue.

Most of the axions with $m_a c^2\gtrsim 10^{-2}$~eV are produced thermally,
and one can show by solving the Boltzmann equation that their combined
present mass density would be 
$\Omega_a\approx (m_a c^2/130 \mbox{ eV})\hub^{-2}$
\cite{Kol90}.  Those in the range 19~eV~$\lesssim m_a c^2 \lesssim$~32~eV
would therefore provide $\Omega_a\approx0.3$, making them potential CDM
candidates.  However, axions decay into photons on timescales proportional 
to $m_a^{-5}$.  With masses as large as this, they would decay rapidly
enough to flood the night sky with ultraviolet light.  Consistency with 
the observed intensity of the extragalactic background light then leads
to the upper bound $m_a c^2\lesssim 3$~eV \cite{Ove00}.

Thermal axions with $10^{-3}$~eV~$\lesssim m_a c^2 \lesssim 3$~eV can be
eliminated on different astrophysical grounds: they couple so weakly to 
ordinary matter that they could stream more or less freely out of the cores
of red giants and supernovae, taking energy with them.
Observations of the neutrino flux from supernova SN1987a show no 
evidence for such an effect \cite{Kol90}.

Axions with $m_a c^2\lesssim 10^{-3}$~eV, finally, are largely 
{\em non\/}thermal, with a collective density given by
$\Omega_a\approx (m_a c^2/4 \mu\mbox{eV})^{-7/6}\hub^{-2}$ 
\cite{Sik00}.\footnote{This is under debate, and may be up to an order of
   magnitude larger if string effects play an important role \cite{Bat00}.}
Rest masses $m_a c^2 \lesssim 10 \mu$eV are then disqualified because
they would provide {\em too much\/} CDM ($\Omega_a \gtrsim 0.5$).  The axion
is thereby restricted to a relatively small window of potential rest masses,
10~$\mu$eV~$\lesssim m_a c^2 \lesssim$~1000~$\mu$eV.  Its plausibility
as a CDM candidate rests on the fact that this allowed window encompasses
the range of values (15~$\mu$eV~$\lesssim m_a c^2 \lesssim$~25~$\mu$eV)
corresponding to $\Omega_a\approx0.3$.
The slow decay rate of axions in this mass range can, moreover, be enhanced 
by trapping them in strong magnetic fields; this is the basis for a number
of ongoing experimental detection efforts in Japan and the 
U.S.A. \cite{Sik00}.

Like axions, supersymmetric WIMPs have their origin in a new symmetry of
nature, spontaneously broken in the early Universe.  This is supersymmetry
(SUSY), which pairs each boson with a fermion superpartner, and vice versa.
Because no such pairs can be formed with the {\em known\/} bosons and 
fermions of the standard model of particle physics, the number of fundamental
particles must be doubled.  The new SUSY partners are presumably so massive
that they have not been discovered yet; the lightest SUSY particle (LSP)
in particular must have a rest mass $m_{\chi} c^2 \gtrsim 50$~GeV \cite{Ell00}.

This LSP plays the role of the WIMP in SUSY theories.  It is stable, 
decaying solely by pair-annihilation with itself (a very slow process).
This is due to an additional new symmetry of nature, known as $R$-parity, 
which is necessary (in ``minimal SUSY'' models) to keep the proton from 
decaying via intermediate SUSY states.
(There are also nonminimal SUSY theories in which this symmetry too is
spontaneously broken, and the proton can decay.)  $R$-parity requires the
number of SUSY (and non-SUSY) partners to be conserved in any reaction,
so that, as the Universe cools, heavier SUSY particles can break down into 
lighter ones, but not into ordinary particles.  Eventually, most of the
SUSY mass in the Universe thus ends up in the form of LSPs.

Using the Boltzmann equation, one can calculate the collective density
$\Omega_{\chi}$ of these particles in terms of a number of free parameters
such as the SUSY-breaking energy scale, and the composition of the
LSP.\footnote{This is most likely to be the neutralino, 
   a linear superposition the photino, the zino and two higgsinos
   (the spin-$1/2$ SUSY partners of the photon, $Z$ and Higgs bosons).
   A less favored candidate, because it annihilates {\em too\/} slowly,
   is the gravitino, the spin-$3/2$ SUSY partner of the graviton
   \cite{Ell84}.}
Because its annihilation cross-section goes as $m_{\chi}^{-2}$, one finds
that an LSP more massive than $\sim3$~TeV would annihilate so slowly as to
overclose the Universe.  This leaves an available SUSY WIMP mass window of
50~GeV~$\lesssim m_{\chi} c^2 \lesssim$~3~TeV.  The collective LSP density
turns out to lie within three orders of magnitude of the critical 
density over most of this range \cite{Tur91}, which makes the SUSY WIMP,
like the axion, a plausible CDM candidate.

Many experiments around the world are currently searching for WIMPs in
this mass range, assuming for example that they are gravitationally bound
in the galactic halo and will occasionally be responsible for scattering
events in target nuclei as the Earth follows the Sun around the Milky Way
\cite{Bau00}.  One team ({\sc Dama}) has claimed evidence for a
$m_{\chi} c^2 = 59^{+17}_{-14}$~GeV WIMP signature at 3$\sigma$ confidence
\cite{Bel00}, but this is disputed by a second ({\sc Cdms}) which
has searched a larger region of parameter space and found nothing
\cite{Abu00}.  Strong constraints have also been placed on WIMPs in 
nonminimal versions of SUSY, where the LSP can decay into photons and 
contribute excessively to the intensity of the diffuse $x$- and
$\gamma$-ray backgrounds \cite{Ove00}.  We watch these
developments with interest.
}

\renewcommand{\baselinestretch}{0.825}
\normalsize
{\footnotesize
\begin{thebibliography}{999}
\bibitem{Abd99} Abdurashitov N et al. (1999) Measurement of the Solar Neutrino
                Capture Rate by SAGE and Implications for Neutrino Oscillations
                in Vacuum. Phys. Rev. Lett. 83: 4686-4689
\bibitem{Abu00} Abusaidi R et al. (2000) Exclusion Limits on the WIMP-Nucleon 
                Cross Section from the Cryogenic Dark Matter Search.
                Phys. Rev. Lett. 84: 5699-5703
\bibitem{Agu00} Aguirre A, Haiman Z (2000) Cosmological Constant or 
                Intergalactic Dust? Constraints from the Cosmic Far-Infrared
                Background. Astrophys. J. 532: 28-36
\bibitem{Alc00} Alcock C et al. (2000) The MACHO Project: Microlensing Results
                from 5.7 Years of LMC Observations. Astrophys. J. 542: 281-307
\bibitem{Ale97} Alexander DR et al. (1997) A Theoretical Approach to Globular
                Cluster Low Main Sequence Stars. Astron. Astrophys. 317: 90-98
\bibitem{Ath98} Athanassopoulos C et al. (1998) Results on 
                $\nu_{\mu} \rightarrow \nu_e$ Neutrino Oscillations from the
                LSND Experiment. Phys. Rev. Lett. 81: 1774-1777
\bibitem{Bah96} Bahcall JN et al. (1996) The Hubble Space Telescope Quasar
                Absorption Line Key Project. VII. Absorption Systems at
                $z_{\scriptscriptstyle {\rm abs}} \leq 1.3$. 
                Astrophys. J. 457: 19-49
\bibitem{Bah98} Bahcall NA, Fan X (1998) The Most Massive Distant Clusters: 
                Determining $\Omega$ and $\sigma_8$. Astrophys. J. 504: 1-6
\bibitem{Bah99} Bahcall NA, Ostriker JP, Perlmutter S, Steinhardt PJ (1999)
                The Cosmic Triangle: Revealing the State of the Universe.
                Science 284: 1481-1488
\bibitem{Bah00} Bahcall NA, Cen R, Dav\'e R, Ostriker JP, Yu Q (2000) The 
                Mass-to-Light Function: Antibias and $\Omega_m$. 
                Astrophys. J. 541: 1-9
\bibitem{Bat00} Battye RA, Shellard EPS (2000) Axion String Cosmology and its
                Controversies.  In: \cite{Kla00}, pp.~565-572
\bibitem{Bau00} Baudis L, Klapdor-Kleingrothaus HV (2000) Direct Detection of
                Nonbaryonic Dark Matter.  In: \cite{Kla00}, pp.~881-898
\bibitem{Bel00} Belli P et al. (2000) Searching for the Dark Universe by the
                DAMA Experiments. In: \cite{Kla00}, pp.~869-880
\bibitem{Ben96} Bennett C et al. (1996) Four-Year COBE DMR Cosmic Microwave
                Background Observations: Maps and Basic Results.
                Astrophys. J. 464: L1-L4
\bibitem{Ber00} Bergmann M, Jorgensen I (2000) Stellar Populations of LSB
                Galaxies: Spectral Observations with the HET.
                Bull. Am. Astron. Soc. 197: \#111.10
\bibitem{Bla99} Blanton M, Cen R, Ostriker JP, Strauss MA (1999) The Physical
                Origin of Scale-Dependent Bias in Cosmological Simulations.
                Astrophys. J. 522: 590-603
\bibitem{Blo85} Blome H-J, Priester W (1985) Vacuum Energy in Cosmic
                Dynamics. Astrophys. Sp. Sci. 117: 327-335
\bibitem{Blo91} Blome H-J, Priester W (1991) Big Bounce in the Very Early
                Universe. Astron. Astrophys. 250: 43-49
\bibitem{Blo97} Blome H-J, Hoell J, Priester W (1997) Kosmologie. In:
                Bergmann-Schaefer: Lehrbuch der Experimentalphysik, vol.~8.
                W.~de~Gruyter, Berlin, pp.~311-427
\bibitem{Car97} Carlberg RG, Yee HKC, Ellingson E (1997) The Average Mass and
                Light Profiles of Galaxy Clusters. Astrophys. J. 478: 462-475
\bibitem{Car99} Carlberg RG et al. (1999) The $\Omega_M$-$\Omega_{\Lambda}$
                Dependence of the Apparent Cluster $\Omega$. 
                Astrophys. J. 516: 552-558
\bibitem{Car01} Carroll SM (2001) The Cosmological Constant. Living Reviews 
                in Relativity {\bf 4}, \url{http://www.livingreviews.org/Articles/Volume4/2001-1carroll}
\bibitem{Cha99} Charbonnel C et al. (1999) Grids of Stellar Models. VIII. From
                0.4 to 1.0 $M_{\odot}$ at $Z=0.020$ and $Z=0.001$, with the
                MHD Equation of State. Astron. Astrophys. Suppl. Ser. 135: 
                405-413
\bibitem{Cle98} Cleveland BT et al. (1998) Measurement of the Solar Electron 
                Neutrino Flux with the Homestake Chlorine Detector.
                Astrophys. J. 496: 505-526
\bibitem{Coh99} Cohen JG, Blandford R, Hogg DW, Pahre MA, Shopbell PL (1999)
                Caltech Faint Galaxy Redshift Survey. VIII. Analysis of the
                Field J0053+1234.  Astrophys. J. 512: 30-47
\bibitem{Coo99} Cooray AR (1999) Cosmological Parameters from Statistics of
                Strongly Lensed Radio Sources. Astron. Astrophys. 342: 353-362
\bibitem{Cow99} Cowan JJ et al. (1999) $r$-Process Abundances and Chronometers
                in Metal-Poor Stars. Astrophys. J. 521: 194-205
\bibitem{Cro99} Croft RAC, Hu W, Dav\'e R (1999) Cosmological Limits on the
                Neutrino Mass from the Ly$\alpha$ Forest. 
                Phys. Rev. Lett. 83: 1092-1095
\bibitem{deB00} de Bernardis P et al. (2000) A Flat Universe from 
                High-Resolution Maps of the Cosmic Microwave Background
                Radiation. Nature 404: 955-959
\bibitem{Din98} Dinshaw N, Foltz CB, Impey CD, Weymann RJ (1998) Ultraviolet
                Spectroscopy of the Quasar Pair LB 9605, LB 9612 with the
                Hubble Space Telescope: Evolution in the Size of the
                Ly$\alpha$ Absorbers? Astrophys. J. 494: 567-580
\bibitem{Dre00} Drell PS, Loredo TJ, Wasserman I (2000) Type Ia Supernovae,
                Evolution, and the Cosmological Constant. Astrophys. J.
                530: 593-617
\bibitem{Efs90} Efstathiou G, Sutherland WJ, Maddox SJ (1990) The Cosmological
                Constant and Cold Dark Matter. Nature 348: 705-707
\bibitem{Ehl89} Ehlers J, Rindler W (1989) A Phase-Space Representation of
                Friedmann-Lema\^{\i}tre Universes Containing Both Dust and
                Radiation and the Inevitability of a Big Bang. 
                Mon. Not. R. Astron. Soc. 238: 503-521
\bibitem{Ell88} Ellis GFR (1988) Does Inflation Necessarily Imply $\Omega=1$?
                Class. Quant. Grav. 5: 891-901
\bibitem{Ell84} Ellis J, Hagelin JS, Nanopoulos DV, Olive K, Srednicki M 
                (1984) Supersymmetric Relics from the Big Bang. 
                Nucl. Phys. B238: 453-476
\bibitem{Ell00} Ellis J, Falk T, Ganis G, Olive KA (2000) Supersymmetric Dark
                Matter in the Light of LEP and the Tevatron Collider.
                Phys. Rev. D62: 075010
\bibitem{Epp00} Eppley JM, Partridge RB (2000) Absence of ``Ghost Images''
                Excludes Large Values of the Cosmological Constant.
                Astrophys. J. 538: 489-492
\bibitem{Fal98} Falco EE, Kochanek CS, Mu\~{n}oz JA (1998) Limits on 
                Cosmological Models from Radio-Selected Gravitational Lenses.
                Astrophys. J. 494: 47-59
\bibitem{Fel93} Feldman HA, Evrard AE (1993) Structure in a Loitering
                Universe. Int. J. Mod. Phys. D2: 113-122
\bibitem{Fel86} Felten JE, Isaacman R (1986) Scale Factors $R(t)$ and
                Critical Values of the Cosmological Constant $\Lambda$.
                Rev. Mod. Phys. 58: 689-698
\bibitem{Fie00} Fields BD, Freese K, Graff DS (2000) Chemical Abundance 
                Constraints on White Dwarfs as Halo Dark Matter. 
                Astrophys. J. 534: 265-276
\bibitem{Fra97} Franx M, Illingworth GD, Kelson DD, van Dokkum PG, Tran K-V
                (1997) Astrophys. J. 486: L75-L78
\bibitem{Fuk00} Fukuda S et al. (2000) Tau Neutrinos Favored over Sterile
                Neutrinos in Atmospheric Muon Neutrino Oscillations.
                Phys. Rev. Lett. 85: 3999-4003
\bibitem{Fuk98b} Fukuda Y et al. (1998) Evidence for Oscillation of
                 Atmospheric Neutrinos. Phys. Rev. Lett. 81: 1562-1567
\bibitem{Fuk90} Fukugita M, Takahara F, Yamashita K, Yoshii Y (1990) 
                Test for the Cosmological Constant with the Number Count
                of Faint Galaxies. Astrophys. J. 361: L1-L4
\bibitem{Fuk98a} Fukugita M, Hogan CJ, Peebles PJE (1998) The Cosmic
                 Baryon Budget. Astrophys. J. 503: 518-530
\bibitem{Gam70} Gamow G (1970) My World Line. Viking Press, New York, p.~44
\bibitem{Gar93} Gardner JP, Cowie LL, Wainscoat RJ (1993) Galaxy Number Counts
                from $K=10$ to $K=23$. Astrophys. J. 415: L9-L12
\bibitem{Gaw98} Gawiser E, Silk J (1998) Extracting Primordial Density
                Fluctuations. Science 280: 1405-1411
\bibitem{Gel89} Geller M, Huchra J (1989) Mapping the Universe.
                Science 246: 897-903
\bibitem{Got89} Gott JR III, Park M-G, Lee HM (1989) Setting Limits on $q_0$
                from Gravitational Lensing. Astrophys. J. 338: 1-12
\bibitem{Gra96} Graff DS, Freese K (1996) The Mass Function of Low-Mass Halo
                Stars: Limits on Baryonic Halo Dark Matter. 
                Astrophys. J. 467: L65-L68
\bibitem{Gri01} Griffiths LM, Melchiorri A, Silk J (2001) Cosmic Microwave 
                Background Constraints on a Baryonic Dark Matter-Dominated
                Universe. Astrophys. J. 553: L5-L8
\bibitem{Gue00} Guerra EJ, Daly RA, Wan L (2000) Global Cosmological 
                Parameters Determined Using Classical Double Radio Galaxies.
                Astrophys. J. 544: 659-670
\bibitem{Gui98} Guinan EF et al. (1998) The Distance to the Large Magellanic
                Cloud from the Eclipsing Binary HV2274. Astrophys. J.
                509: L21-L24
\bibitem{Ham99} Hampel W et al. (1999) GALLEX Solar Neutrino Observations:
                Results for GALLEX IV. Phys. Lett. B447: 127-133
\bibitem{Han00} Hanany S et al. (2000) MAXIMA-1: A Measurement of the Cosmic 
                Microwave Background Anisotropy on Angular Scales of 
                $10^{\prime}-5^{\circ}$. Astrophys. J. 545: L5-L9
\bibitem{Han99} Hansen BMS (1999) The Origin of Primordial Dwarf Stars and
                Baryonic Dark Matter. Astrophys. J. 517: L39-L42
\bibitem{Heg86} Hegyi DJ, Olive KA (1986) A Case Against Baryons in Galactic
                Halos. Astrophys. J. 303: 56-65
\bibitem{Her99} Herrnstein JR et al. (1999) A Geometric Distance to the
                Galaxy NGC4258 from Orbital Motions in a Nuclear Gas Disk.
                Nature 400: 539-541
\bibitem{Hil00} Hillebrandt W, Niemeyer JC (2000) Type Ia Supernova 
                Explosion Models. Ann. Rev. Astron. Astrophys. 38: 191-230
\bibitem{Hoe94} Hoell J, Liebscher D-E, Priester W (1994) Confirmation of
                the Friedmann-Lema\^{\i}tre Universe by the Distribution of
                the Larger Absorbing Clouds. Astron. Nachr. 315: 89-96
\bibitem{Hog00} Hogan CJ, Dalcanton JJ (2000) New Dark Matter Physics: Clues
                from Halo Structure. Phys. Rev. D62: 063511
\bibitem{Hu94} Hu EM, Ridgway SE (1994) Two Extremely Red Galaxies. 
               Astron. J. 107: 1303-1306
\bibitem{Hu00} Hu W, Barkana R, Gruzinov A (2000) Fuzzy Cold Dark Matter:
               The Wave Properties of Ultralight Particles. 
               Phys. Rev. Lett. 85: 1158-1161
\bibitem{Hub91} H\"ubner P, Ehlers J (1991) Inflation in Curved Model
                Universes with Noncritical Density. 
                Class. Quant. Grav. 8: 333-346
\bibitem{Iba99} Ibata RA, Richer HB, Gilliland RL, Scott D (1999) Faint,
                Moving Objects in the Hubble Deep Field: Components of the
                Dark Halo? Astrophys. J. 524: L95-L97
\bibitem{Jaf00} Jaffe AH et al. (2000) Cosmology from MAXIMA, BOOMERANG
                \& COBE/DMR CMB Observations. Phys. Rev. Lett. 86: 3475-3479
\bibitem{Jen98} Jenkins A et al. (1998) Evolution of Structure in Cold Dark
                Matter Universes. Astrophys. J. 499: 20-40
\bibitem{Kin01} Kinney WH, Brisudova M (2001) An Attempt to Do Without Dark
                Matter. In: Fry JN, Buchler JR and Kandrup H (eds)
                The Onset of Nonlinearity in Cosmology
\bibitem{Kir97} Kirkman D, Tytler D (1997) Intrinsic Properties of the 
                $<z>=2.7$ Ly$\alpha$ Forest from Keck Spectra of Quasar
                HS 1946+7658. Astrophys. J. 484: 672-694
\bibitem{Kla99} Klapdor-Kleingrothaus HV, Baudis L (1999)
                Dark Matter in Astrophysics and Particle Physics 1998.
                Institute of Physics Press, Oxford
\bibitem{Kla00} Klapdor-Kleingrothaus HV, Krivosheina IV (2000)
                Beyond the Desert 1999. Institute of Physics Press, Oxford
\bibitem{Kly99} Klypin AA, Kravtsov AV, Valenzuela O, Prada F (1999)
                Where are the Missing Galactic Satellites?
                Astrophys. J. 522: 82-92
\bibitem{Koc96a} Kochanek CS (1996) The Mass of the Milky Way.
                 Astrophys. J. 457: 228-243
\bibitem{Koc96b} Kochanek CS (1996) Is There a Cosmological Constant?
                 Astrophys. J. 466: 638-659
\bibitem{Kol90} Kolb EW, Turner MS (1990) The Early Universe.
                Addison-Wesley, Reading
\bibitem{Kra00} Krauss L (2000) Quintessence: The Mystery of the Missing
                Mass in the Universe. Basic Books, New York
\bibitem{Lie92a} Liebscher D-E, Priester W, Hoell J (1992) A New Method to
                 Test the Model of the Universe.
                 Astron. Astrophys. 261: 377-381
\bibitem{Lie92b} Liebscher D-E, Priester W, Hoell J (1992) Lyman$\alpha$
                 Forests and the Evolution of the Universe.
                 Astron. Nachr. 313: 265-273
\bibitem{Lu96} Lu L, Sargent WLW, Womble DS, Takada-Hidai M (1996) The 
               Ly$\alpha$ Forest at $z\sim4$: Keck HIRES Observations of
               Q0000-26. Astrophys. J. 472: 509-531
\bibitem{Lum98} Luminet, J-P (1998) Past and Future of Cosmic Topology.
                Acta Cosmologica 24: 105-115
\bibitem{Mal97} Malhotra S, Rhoads JE, Turner EL (1997) Through a Lens Darkly:
                Evidence for Dusty Gravitational Lenses.
                Mon. Not. R. Astron. Soc. 288: 138-144
\bibitem{Mao99} Maoz E et al. (1999) A Distance to the Galaxy NGC4258 from
                Observations of Cepheid Variable Stars. Nature 401: 351-354
\bibitem{McG01} McGaugh SS (2001) Dynamics and the Second Peak: Cold Dark
                Matter?  Int. J. Mod. Phys. A.
\bibitem{Mir96} Miralda-Escud\'e J, Cen R, Ostriker JP, Rauch M (1996) 
                The Ly$\alpha$ Forest from Gravitational Collapse in the 
                Cold Dark Matter + $\Lambda$ Model. 
                Astrophys. J. 471: 582-616
\bibitem{Moo94} Moore B (1994) Evidence against Dissipationless Dark Matter 
                from Observations of Galaxy Haloes. Nature 370: 629-631
\bibitem{Mou00} Mould JR et al. (2000) The Hubble Space Telescope Key Project
                on the Extragalactic Distance Scale. XXVIII. Combining the
                Constraints on the Hubble Constant. Astrophys. J. 529: 786-794
\bibitem{Nel00} Nelson CA et al. (2000) Halo Lensing or Self-lensing?
                Locating the MACHO lenses.
                Bull. Am. Astron. Soc. 197: \#04.19
\bibitem{Oli00} Olive KA (2000) Big Bang Nucleosynthesis. Nucl. Phys. Proc. 
                Suppl. 80: 79-93
\bibitem{Ove98} Overduin JM, Cooperstock FI (1998) Evolution of the Scale
                Factor with a Variable Cosmological Term. Phys. Rev. D58:
                043506
\bibitem{Ove00} Overduin JM, Wesson PS (2000) Observational Limits on 
                Nonstandard Dark Matter Candidates. In \cite{Kla00}, 
                pp.~539-544
\bibitem{Pad93} Padmanabhan T (1993) Structure Formation in the 
                Universe. Cambridge Univ. Press., Cambridge, p.~63
\bibitem{Pee93} Peebles PJE (1993) Principles of Physical Cosmology.
                Princeton Univ. Press., Princeton
\bibitem{Pee00a} Peebles PJE (2000) Fluid Dark Matter. 
                 Astrophys. J. 534: L127-L129
\bibitem{Pee00b} Peebles PJE, Seager S, Hu W (2000) Delayed Recombination.
                 Astrophys. J. 539: L1-L4
\bibitem{Per99} Perlmutter S et al. (1999) Measurements of $\Omega$ and
                $\Lambda$ from 42 High-Redshift Supernovae. 
                Astrophys. J. 517: 565-586
\bibitem{Pri95} Priester W, Hoell J, Blome, H-J (1995) The Scale of the
                Universe: A Unit of Length.  Comments Astrophys. 17: 327-342
\bibitem{Pri98} Priester W, van de Bruck C (1998) 75 Jahre Theorie des
                expandieren Kosmos: Friedmann Modelle und der
                ``Einstein-Limit.'' Naturwiss. 85: 524-538
\bibitem{Pri00} Primack JR (2000) Status of Cosmology.  In:
                Courteau S and Willick J (eds) Cosmic Flows:
                Towards an Understanding of Large-Scale Structure,
                ASP Conference Series, vol.~201
\bibitem{Rau98} Rauch M (1998) The Lyman Alpha Forest in the Spectra of
                Quasistellar Objects. Ann. Rev. Astron. Astrophys. 36: 267-316
\bibitem{Rie98} Riess AG et al. (1998) Observational Evidence from Supernovae 
                for an Accelerating Universe and a Cosmological Constant.
                Astron. J. 116: 1009-1038
\bibitem{Rin77} Rindler W (1977) Essential Relativity, 2nd edn 
                Springer-Verlag, Berlin, \S9.9.
\bibitem{Rio91} Riordan M, Schramm DN (1991) The Shadows of Creation.
                Freeman, New York
\bibitem{Roc00} Rocca-Volmerange B (2000) Constraints on the Cosmological 
                Constant $\Omega_{\Lambda}$ from Faint Galaxy Counts. In: 
                Mazure A, Le F\`evre O, Le Brun V (eds) Clustering at High 
                Redshift, ASP Conference Series, vol.~200, pp.~163-167
\bibitem{Sah92} Sahni V, Feldman H, Stebbins A (1992) Loitering Universe.
                Astrophys. J. 385: 1-8
\bibitem{Sch93} Schneider P (1993) Upper Bounds on the Cosmological Density of
                Compact Objects with Sub-solar Masses from the Variability of
                QSOs. Astron. Astrophys. 279: 1-20
\bibitem{Sci93} Sciama DW (1993) Modern Cosmology and the Dark Matter Problem.
                Cambridge Univ. Press, Cambridge
\bibitem{Sha01} Shanks T et al. (2001) Evidence for Galaxy Formation at
                High Redshift.  In: The Extragalactic Infrared Background
                and its Cosmological Implications, IAU Symposium 204
\bibitem{Sik00} Sikivie P (2000) Axions and Dark Matter Caustics. 
                In: \cite{Kla00}, pp.~547-564
\bibitem{Spe00} Spergel DN, Steinhardt PJ (2000) Observational Evidence for
                Self-Interacting Cold Dark Matter.
                Phys. Rev. Lett. 84: 3760-3763
\bibitem{Sta66} Stabell R, Refsdal S (1966) Classification of General 
                Relativistic World Models. 
                Mon. Not. R. Astron. Soc. 132: 379-388
\bibitem{Teg00} Tegmark M, Zaldarriaga M (2000) New CMB Constraints on the
                Cosmic Matter Budget: Trouble for Nucleosynthesis?
                Phys. Rev. Lett. 85: 2240-2243
\bibitem{Tot97} Totani T, Yoshii Y, Sato K (1997) Evolution of the Luminosity
                Density in the Universe: Implications for the Nonzero 
                Cosmological Constant. Astrophys. J. 483: L75-L78
\bibitem{Tot00} Totani T, Yoshii Y (2000) Unavoidable Selection Effects in the
                Analysis of Faint Galaxies in the Hubble Deep Field: Probing
                the Cosmology and Merger History of Galaxies. 
                Astrophys. J. 540: 81-98
\bibitem{Tri00} Tripp TM, Savage BD, Jenkins EB (2000) Intervening O {\sc vi}
                Quasar Absorption Systems at Low Redshift: A Significant 
                Baryon Reservoir. Astrophys. J. 534: L1-L5
\bibitem{Tur91} Turner MS (1991) Dark Matter in the Universe.
                Physica Scripta T36: 167-182
\bibitem{Tyt00} Tytler D, O'Meara JM, Suzuki N, Lubin D (2000) Review of 
                Big Bang Nucleosynthesis and Primordial Abundances. 
                Physica Scripta 85: 12
\bibitem{Uda00} Udalski A (2000) The Optical Gravitational Lensing Experiment:
                Red Clump Stars as a Distance Indicator.  Astrophys. J.
                531: L25-L28
\bibitem{van99} van de Bruck C, Priester W (1999) The Cosmological Constant
                $\Lambda$, the Age of the Universe and Dark Matter: Clues from
                the Ly$\alpha$-Forest.  In: \cite{Kla99}, pp.~181-196
\bibitem{Wei99} Weinberg DH, Croft RAC, Hernquist L, Katz N, Pettini M (1999)
                Closing in on $\Omega_M$: The Amplitude of Mass Fluctuations
                from Galaxy Clusters and the Ly$\alpha$ Forest. 
                Astrophys. J. 522: 563-568
\bibitem{Wei01} Weinberg S (2001) The Cosmological Constant Problems.
                In Cline D (ed) Sources and Detection of Dark Matter and
                Dark Energy in the Universe, Springer, pp.~18-26
\bibitem{Whi96} White M, Scott D (1996) Why Not Consider Closed Universes?
                Astrophys. J. 459: 415-431
\bibitem{Whi00} White M, Scott D, Pierpaoli E (2000) Boomerang Returns
                Unexpectedly. Astrophys. J. 545: 1-5
\bibitem{Whi93} White SDM, Navarro JF, Evrard A, Frenk CS (1993) The Baryon
                Content of Galaxy Clusters: a Challenge to Cosmological
                Orthodoxy. Nature 366: 429-433
\bibitem{Wil00a} Willick JA, Batra P (2000) A Determination of the Hubble
                 Constant from Cepheid Distances and a Model of the Local
                 Peculiar Velocity Field. Astrophys. J. 548: 564-584
\bibitem{Wil00b} Williger GM, Smette A, Hazard C, Baldwin JA, McMahon RG 
                 (2000) Evidence for Large-Scale Structure in the Ly$\alpha$
                 Forest at $z > 2.6$. Astrophys. J. 532: 77-87
\bibitem{Zeh99} Zehavi I, Dekel A (1999) Evidence for a Positive Cosmological
                Constant from Flows of Galaxies and Distant Supernovae.
                Nature 401: 252-254
\end{thebibliography}
}
\end{document}